\documentclass[a4paper,12pt]{article}

\usepackage[pdftex,breaklinks,colorlinks,pagebackref=true]{}

\usepackage{amstext,amsmath,amssymb}        
\usepackage{txfonts}                

\usepackage[usenames,dvipsnames]{color}  
\usepackage{colortbl}    
\usepackage{ textcomp }     
\usepackage{sidecap}

\usepackage{fancyhdr}               
\usepackage{vmargin}
\usepackage{appendix}
\usepackage{makeidx}
\usepackage{color}
\usepackage{url}

\usepackage{remreset}

\usepackage{graphicx}

\usepackage{wrapfig}
\newlength{\wrapsep}
\newlength{\saveintextsep}
\newlength{\wrapsepcol}
\newlength{\savecolumnsep}
\usepackage{epsfig}
\usepackage{rotating}
\usepackage{subfigure}

\usepackage[square,comma,numbers,sort&compress]{natbib}
\usepackage{hypernat}  

\usepackage[spanish,english]{babel}
\usepackage{graphicx} 
\usepackage{comment} 
\usepackage{url}
\usepackage{bm}
\usepackage{algorithmic}
\usepackage{ifthen}   
\usepackage{amsgen} 
\usepackage{xkeyval}

\usepackage{marvosym}
\usepackage{upgreek}
\usepackage{sidecap}

\usepackage[T1]{fontenc}
\usepackage[ansinew]{inputenc}

\makeatletter
  \def\cleardoublepage{\clearpage\if@twoside \ifodd\c@page\else%
    \hbox{}
    \thispagestyle{empty}               
       \newpage
    \if@twocolumn\hbox{}\newpage\fi\fi\fi}
\makeatother

\begin{document}


\begin{center}
\LARGE{\bfseries {A review of High Performance Computing foundations for scientists}}
\end{center}

\begin{center}
\large{Pablo Garc\'ia-Risue\~no} \\[10pt]
\small{Institut f\"ur Physik, Humboldt Universit\"at zu Berlin, 12489 Berlin, Germany} \\
\vspace{0.5cm}
\large{Pablo E. Ib\'a\~nez} \\[10pt]
\small{Departamento de Inform\'atica e Ingenier\'ia de Sistemas, Universidad de Zaragoza -} \\
\small{Instituto de Investigaci\'on en Ingenier\'ia de Arag\'on (I3A), E-50018 SPAIN}
\end{center}

\abstract{The increase of existing\index{work of a computational scientist|(} computational capabilities
has made simulation emerge as a third discipline 
of Science, lying midway between experimental and purely theoretical branches \cite{makovsimulation,Cra2002Book}.
Simulation enables the evaluation of quantities which otherwise would not be accessible, helps to 
improve experiments and provides new insights on systems which are analysed 
\cite{germann_Pflop_MD,PhysRevLett.90.258101,Shaw15102010,CPHC:CPHC200600128}.
Knowing the fundamentals of computation can be very useful for scientists, for it can help them to
improve the performance of their theoretical models and simulations. This review includes some 
technical essentials that can be useful to this end, and it is devised as a complement for researchers
whose education is focused on scientific issues and not on technological respects. In this 
document we attempt to discuss the fundamentals of High Performance Computing (HPC) \cite{libro_HPC}
in a way which
is easy to understand without much previous background. We sketch the way standard computers and 
supercomputers work, as well as discuss distributed computing and discuss essential aspects to take into account 
when running scientific calculations in computers.
}

\vspace{0.5cm}


\vspace{0.2cm}

\section{Introduction}

\index{computation|(}
Scientific computer simulation is a very useful research tool. 
Its usefulness can be classified into (at least) the following three situations \cite{Cra2002Book}
\begin{itemize}
\item If an experiment reproducing the simulated physical or chemical process
is not carried out: Sometimes, one wants to investigate a phenomenon, and making an 
experiment for it is too costly, expensive, dangerous, slow, or presents other 
inconveniences which advise against carrying it out \cite{germann_Pflop_MD}.
In other cases, it simply cannot be conducted due to the 
lack of the appropriate conditions or technology.
In these cases, simulation can generate the sought information. 
\item If an experiment reproducing the physical or chemical simulated process is 
carried out, but it is not completely understood: Simulations can also be run in order to 
explain phenomena which arise in performed experiments, but whose explanation is unclear. With simulation, 
special conditions can be selected, enabling a focus on arbitrarily chosen aspects.
\item If an experiment reproducing the physical or chemical simulated process
is carried out, but it can be improved: The conditions for an experiment can be chosen
among an ample collection of possibilities. If experiment and simulation have an information feedback, 
the former can be tuned to provide researchers with more accurate results. A simulation can also 
be performed prior to the corresponding experiment, with the goal of obtaining information useful in
order to choose which specific experiment to do.
\end{itemize}

Simulation is suitable to analyse a wide range of systems. It is especially important for studying 
atomic and molecular systems because
many physical and chemical features of the systems emerge from small scale phenomena, which frequently hinders
direct experimentation and makes \emph{in silico} techniques advisable or compulsory.

In order to provide reliable and non-trivial results, simulation must be carried out
respecting some conditions. On the one hand, the theory that the calculations are based on
must be appropriate for the simulated system. On the other hand, the simulation has to be able to
provide accurate results in a reasonable time. The last point is often a bottleneck for simulations; 
for example, solving some quantum equations for a system of 
thousands of atoms could even take years in small computers. 

In the last decades, computation technologies have improved enormously,
and currently, even the computers that can 
be found in a common computer shop are capable of performing tens of billions ($10^{10}$) of 
operations, such as additions or products,
per second. For scientific purposes there exist special machines 
which greatly increase this performance. The field which studies the way such machines work and their
use in specific problems is called High Performance Computing (HPC) or supercomputing. The limit 
between HPC and ordinary computing is rather arbitrary, since HPC refers merely to calculations
that are run in computers which are more powerful than standard ones.
The enhanced calculation capabilities of supercomputers are to a great extent achieved both by doing every operation 
in a extremely brief time and by having many computing units performing operations simultaneously. The latter
feature is called \emph{parallelism},\index{parallelism} and it is essential for High Performance Computing
(HPC). Parallelism is a challenge both for computer 
architecture designers and for software developers\index{software developer}, 
since data must be transferred throughout
the parts of the computer in an efficient way. 
To this end, specific programming
interfaces exist, among which we can highlight MPI and OpenMP \cite{MPI_OpenMP_HPC}.
It is worth stressing that running a parallelizable program in $N_c$ computing units hardly ever makes
its execution $N_c$ times faster than running it in just one. 
This is mainly due to two reasons. First, 
most algorithms require data transfer among computing units.
This slows the calculations down because the devices for transferring information among
computing units in a computer do work more slowly than the computing units themselves 
(see sec. \ref{Clusters}). Second, algorithms typically
cannot be parallelized in a complete way. 
Only one part of an algorithm can be shared among several computing units, 
while the rest of it is inherently serial, or it has to be executed in a number of cores lower than the 
available number of cores (see Amdahl's Law, later discussed). 
Despite these problems,
high performance machines have acquired major importance in present science.

Having in mind some fundamentals of how computers work can be useful for scientists because it can help
them to know when simulations could boost their research and in which way simulations should be conducted.
Having this knowledge can help, for example, to choose the appropriate machine for the calculations,
to choose the theory level for the simulations, or to understand the source of the errors in the simulations
and how these could be overcome. Since simulation has become a very important tool for present-day 
science, we consider some basics regarding simulation should be known by scientists.

This paper is structured as follows. In section \ref{hardwarebasics} we discuss the essentials of 
a computer. In section \ref{btvnp}, we point out the strategies that have been followed to greatly increase
the performance of computers. In section \ref{Clusters} we focus on standard designs of high performance
computers, and in section \ref{hahm} we sketch other useful schemes. In section \ref{ditribcomp} we discuss
some fundamentals of distributed computing, which is a powerful alternative to traditional 
(physically localized) computing machines. In section \ref{workcompsci} we make some general remarks on 
basic limitations of the two aspects which are customarily most important in computations: accuracy and efficiency.

\section{Hardware basics}\label{hardwarebasics}
\begin{figure}[!ht]
\begin{center}
\includegraphics[width=3.7in]{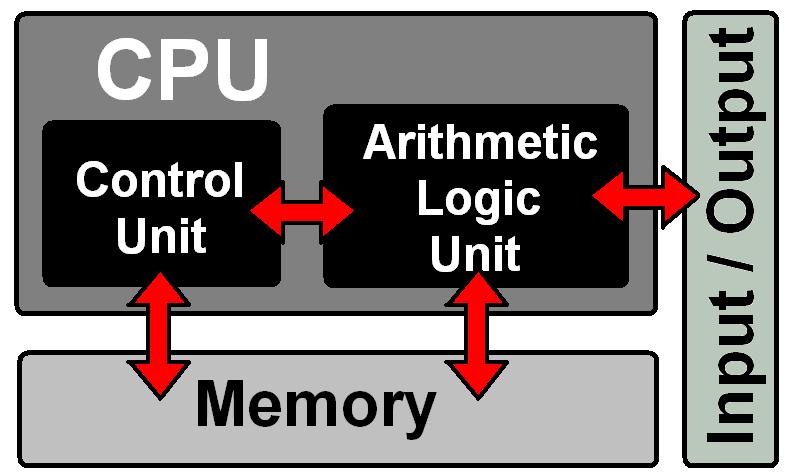}
\caption{\label{fig:comp1} {\footnotesize
Scheme of the basic parts of a computer \cite{libro_HPC}.
The red arrows indicate information flow.
}}
\end{center}
\end{figure}

The \index{basic computer} basic scheme of a single computer 
is simple; it is sketched in fig. \ref{fig:comp1}. A computer contains \emph{memory}, which is
a set of physical devices to store information. The memory contains
both data to use in the calculations, and coded instructions to be provided to the \emph{Control Unit} 
so that the calculations can be performed. The Control Unit 
controls the data flow and the operations that are to be performed in the Arithmetic Logic Unit
(ALU).
The ALU performs both arithmetic operations on numbers (like addition and subtraction) and
logic operations (like AND, OR, XOR, etc.) on the binary digits (bits) of the stored variables.
Computers also include a \emph{clock}, which operates at a given frequency (the clock frequency or 
clock-rate\index{clock}\index{clock!frequency}\index{clock!rate}). The 
clock-rate\index{clock!frequency}\index{clock!rate} determines the number of maximum 
operations performed per second: 
An arithmetic or logic operation (as well as each stage an operation is divided into) takes at least
one clock cycle. 
The Control Unit and the Arithmetic Logic Unit together form the 
CPU. The basic computer device also includes an 
interface which enables its interaction with the human user (input/output).
High performance computers are essentially formed by the accumulation of CPUs linked in a smart way,
as we will see later.\index{CPU}

In fig. \ref{fig:comp1}, red arrows indicate information flow. This flow can be physically handled
by different devices (the network)\index{network}. One or several\index{CPU}
CPUs together with some communication 
devices can be set on a thin layer semiconductor\index{semiconductor}
with electronic circuits (i.e., on a chip\index{chip}), to form a processor\index{processor}
or microprocessor\index{processor}. The CPU interacts with the outside world via the \emph{input/output}
interface. This interface enables, for example, the system to be managed by the human user (e.g., 
a keyboard is an input device, and a monitor is an output device).

\begin{SCfigure}
  \centering
  \includegraphics[width=0.365\textwidth]%
    {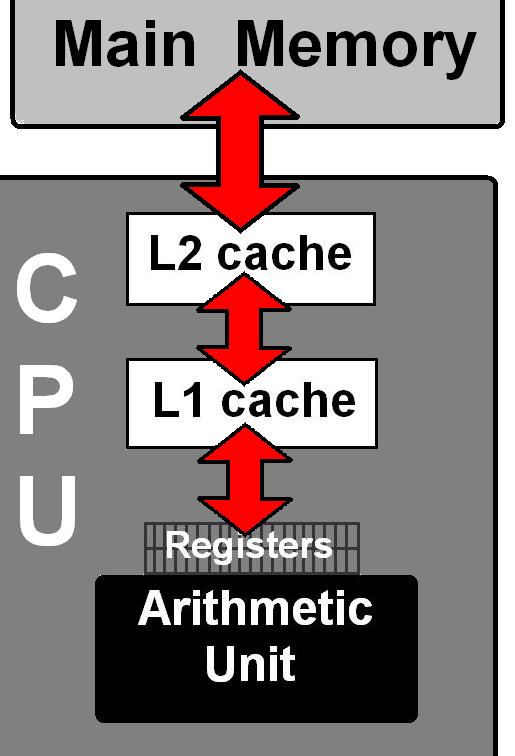}
  \caption{\label{fig:comp2} \footnotesize Scheme of the flow of information (red arrows) in a computer 
to and from its Arithmetic Unit through a hierarchy of 
memories\index{hierarchy of memories}\index{memory!hierarchy} \cite{libro_HPC}. The cache can 
have more levels (not necessarily two). Usually, the closer a device is to the Arithmetic Unit, the 
faster the information transferring, 
but the lower the storage capacity (see fig. \ref{fig:memories}). The disk, as well as eventual 
external memories\index{external memories}\index{memory!non-local} like tape libraries,
are not directly connected to the CPU.
Some architectures also include
direct access paths which connect the registers with the main memory.}
\end{SCfigure}

\begin{figure}[h]
\begin{center}
\includegraphics[width=5.9in]{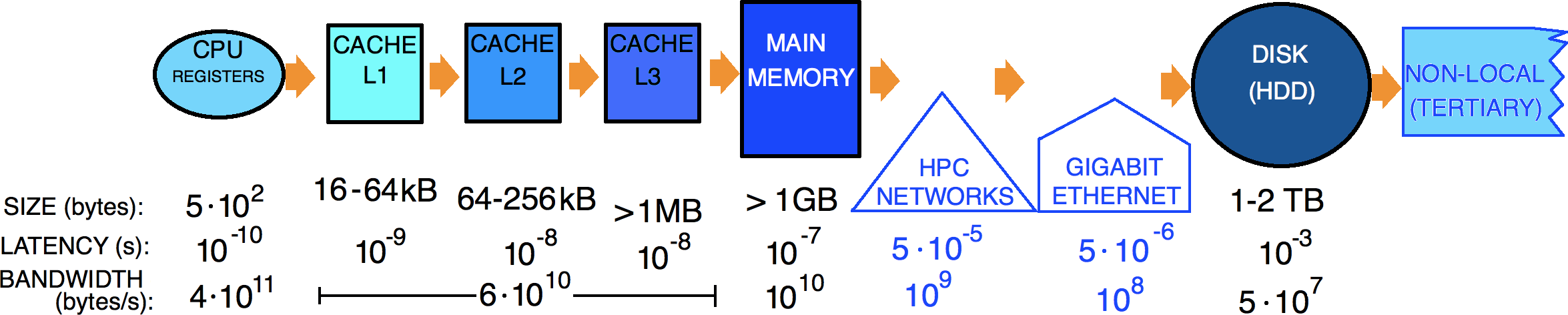}
\caption{\label{fig:memories} {\footnotesize
Hierarchy of memories in a computer \cite{libro_HPC,HenYYYYBook2012,microprocessoreport},\index{hierarchy of memories}
with typical values of their latency\index{latency}, memory size and bandwidth\index{bandwidth}.
The lower the delay in data transfer (latency\index{latency}), the 
higher\index{HPC networks}\index{network!HPC networks}
the bandwidth\index{bandwidth} and the\index{Gigabit ethernet}\index{network!Gigabit ethernet} 
lower the size of the memory. Blue borders indicate elements typical in supercomputers\index{supercomputer},
while black borders indicate elements which are present both in supercomputers and in personal computers.
White shapes do not represent memories, but networking devices 
for interconnection of nodes in supercomputers\index{network}\index{supercomputer}.
}}
\end{center}
\end{figure}

The data moves to and from the ALU through a memory 
hierarchy\index{hierarchy of memories}\index{memory!hierarchy}
following the pattern displayed in fig. \ref{fig:comp2}. 
Information can be stored in numerous devices which exist for that purpose (the memory\index{memory}),
each having a different 
maximum amount of bytes for storage (size) and different bandwidth\index{bandwidth}\index{bandwidth}
(maximum rate for information transfer). 
They also have distinct latencies\index{latency} (the latency\index{latency} is the amount of time between
a request to the memory, and the time when its reply takes place). 
For example, if $y$ bytes of information are to be transferred from a memory device which has a latency\index{latency}
of $l$ seconds, and a bandwidth\index{bandwidth} of $b$ bytes/second, then the minimal
time required for the information to be delivered will be 
$t=l+y/b$. Not only do the different kinds of memory have a latency\index{latency} and a bandwidth\index{bandwidth}, 
but also the network does.
Latencies and bandwidths\index{bandwidth} have a major influence on a computer's performance,
especially in parallel machines (see section \ref{Clusters}).
In fig. \ref{fig:comp2} we can see the scheme of connection of an arithmetic unit
with several types of memories.
Commonly, closer connections are with memories with lower latency\index{latency} 
and higher bandwidth\index{bandwidth}, 
though smaller size. A scheme of the
memory hierarchy\index{hierarchy of memories}, including their present-day typical sizes,
bandwidths\index{bandwidth} and latencies\index{latency}, is displayed in fig. \ref{fig:memories}.

The information which is expected 
to be used immediately by the CPU\index{CPU} is stored in its registers\index{memory!registers}, which 
have a very low latency\index{latency}, but can only store a small amount of information. At present,
typical CPUs\index{CPU} have between 16
and 128 user-visible registers\index{registers}\index{memory!registers} \cite{libro_HPC}. 
The next level in the hierarchy of memories (after the registers) is the level of  
cache\index{cache} memories. Typically, there exist three different levels within this cache memory, which are usually
denoted with L1, L2, L3. 
When the CPU\index{CPU} needs some data, it first checks whether or not they are stored in the cache\index{memory ! cache}. 
Efforts in circuit integration are specifically aimed at increasing the storage capabilities of caches, 
in order to reduce the time\index{time!to access memory} to access the information the CPU\index{CPU} requires. 

After caches, the next level in the memory hierarchy is the main 
memory (which is sometimes called RAM, although this acronym refers to a specific type of 
technology\index{main memory!RAM}\index{memory!RAM}). 

The \emph{disk}\index{disk}\index{memory!disk} (hard disk drive, or HDD) has larger storage space, 
but higher latency\index{latency} and lower bandwidth\index{bandwidth}.
Although access to the disk is the slowest if compared to access to other memories, 
it has much more capability of storing information permanently 
\cite{PARALLEL_HI-PERFORMANCE_COMPUT_CHEMISTRY}
(external memories\index{external memories}\index{memory!non-local}, such as CD-ROMs, pen drives, 
etc. excepted).
The first disk\index{disk}\index{memory!disk} memory was developed by IBM in 1956. This first device was able 
to store  2 kilobits/in\textsuperscript{2}, while 
disks\index{disk}\index{memory!disk} manufactured today can store data at densities of 0.25 
Terabits/in\textsuperscript{2} \cite{memories}. In the last decades, the space of data volumes
is doubling each year or even faster 
\cite{data_management_2005,expendingmooresdividend}.
It is worth mentioning that the increase in disk\index{disk}\index{memory!disk} memory capabilities has been boosted 
by the discovery of giant magnetoresistance \cite{fertmagnetoresistance_original, magnetoresistance_2}. This 
phenomenon makes it possible to manufacture MRAM memories\index{MRAM memory} which store information
(bits) in magnetic layers \cite{sundongmram}, resulting in storage capabilities larger than those of
previous technologies.

The increase in memory size of devices such as caches, RAMs and disks
is quite useful for scientific simulation, because much of the information of the tackled
complex systems has to be stored frequently during the calculation process, which makes 
memory an 
important limiting factor for \emph{in silico} scientific calculations. Both the amount of available 
memory (in disk) and the speed to access information in all levels of the hierarchy imply major
limitations to scientific calculations.
Data storage is reported to be a big energy consumer; moreover, its power intake tends to grow because
storage requirements are increasing over and over, and disks are faster and faster \cite{storage_1} .
The low speed to 
access the information on disks is another drawback of the current technology.
I/O (input and output to disk) bandwidth has not advanced as much as
storage capacity. As stated in \cite{data_management_2005}:
'In the decade 1995-2005, while capacity has grown more than 100-fold, storage bandwidth\index{bandwidth} has 
improved only about 10-fold'.

\begin{figure}[!ht]
\begin{center}
\includegraphics[width=2.6in]{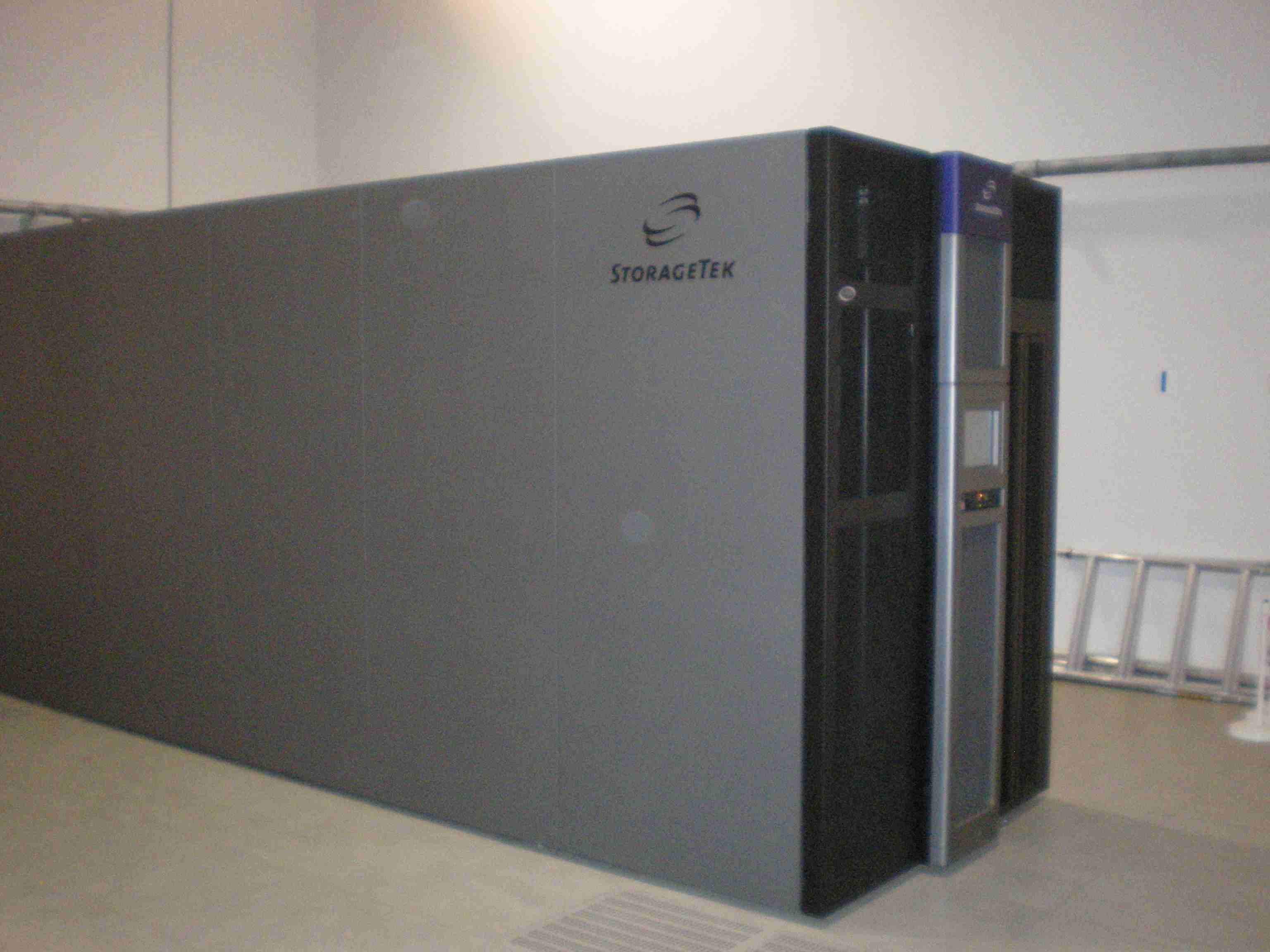}
\caption{\label{fig:externalmemo} {\footnotesize Non-local tape memory at the JSC (J\"ulich, Germany).
}}
\end{center}
\end{figure}

External\footnote{Sometimes called \emph{non-local}\index{memory!non-local}.} 
devices to store information\index{external memories} can be considered the last level in
the hierarchy of memories. 
These devices can be CD or DVD disks, USB flash drives, or different technologies.
Massive storage devices, such as tape libraries (see fig. \ref{fig:externalmemo}),
are often used in supercomputers.
Sometimes, the words \emph{primary memory}\index{memory!primary}
for registers, cache and
main memory, \emph{secondary memory}\index{memory!secondary} for hard disks and
\emph{tertiary memory}\index{memory!tertiary} for non-local memories are used.

\section{Beyond the Von Neumann paradigm}\label{btvnp}
\begin{figure}[!ht]
\begin{center}
\includegraphics[width=5.5in]{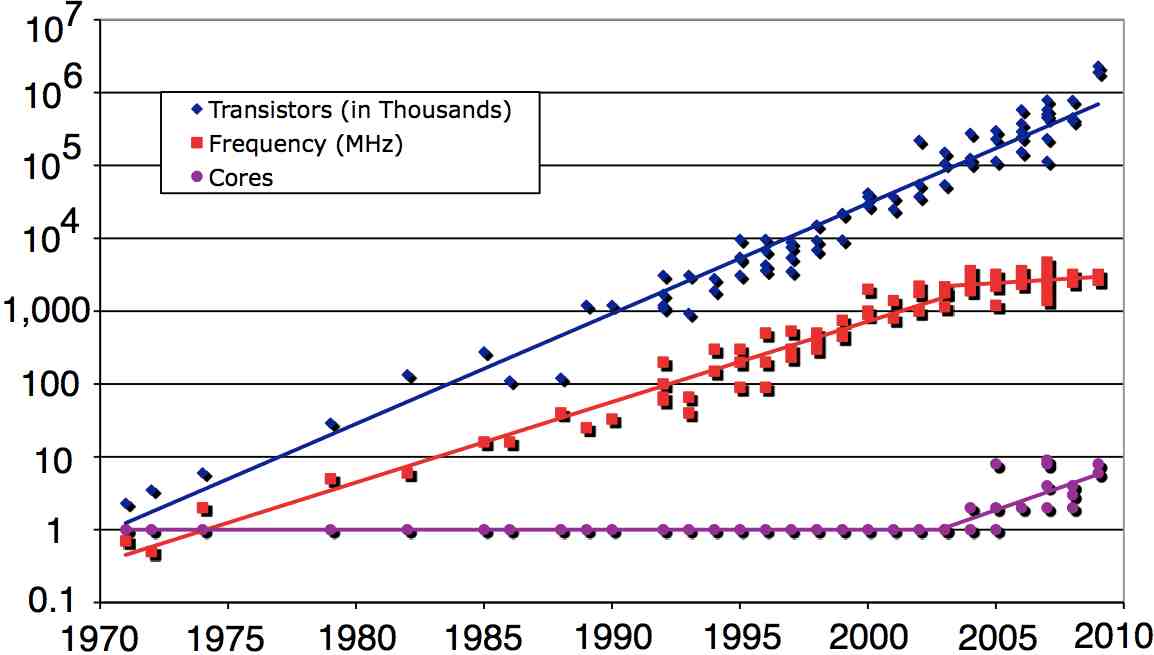}
\caption{\label{fig:integration_evolution} {\footnotesize
Evolution in time of the number of transistors per chip (blue), clock-rate (red) \index{evolution of computers}
and number of core\index{core}s per processor\index{processor} (purple).
It can be appreciated that during the last years the clock-rate\index{clock!frequency}\index{clock!rate} has collapsed, while multicore\index{core} solutions
have started to be implemented. The number of transistors continues to grow exponentially at a steady rate, 
satisfying Moore's Law\index{Moore's Law} to be still satisfied
(Courtesy J. Dongarra\index{Jack Dongarra}; data from \cite{future_of_microprocessors}).
}}
\end{center}
\end{figure}

The basic computer scheme of fig. \ref{fig:comp1} comes from the Von Neumann paradigm \cite{vonneumanoriginal},
which was defined in the 
early times of computer architecture. This scheme has been kept across time because new designs have been required
to work with previous software, which 
(in the beginning) was devised to be coherent with Von Neumann's pattern. 
However, this architecture is not optimal for many present-day purposes, since it is inherently sequential
\cite{future_of_microprocessors}. The CPU\index{CPU} can only deal with one instruction with (possibly) a single 
operand or a group of operands at a time (i.e., at a given clock cycle).
Another main drawback is the \emph{Von Neumann bottleneck}\index{bottleneck}, which results from the fact that
data and instructions must be continually provided to the CPU,\index{CPU} thus making it inevitable 
to perform many time-consuming
accesses to the memory.
Attempts to improve the speed of performing operations continue to be developed. Some of them are related
to software techniques and compilers, while others are related to the CPU\index{CPU} machinery. Among the latter, 
we highlight the following ones \cite{libro_HPC}:
\begin{itemize}
\item  {\bf Integration}: Semiconductors\index{semiconductor} research has boosted the miniaturization of transistors, 
thus enabling an exponential growth of the number of them which can be included on a single chip
(see fig. \ref{fig:integration_evolution}). Since the 
1960s, the number of transistors that can be included on a semiconductor chip approximately doubles every two years.
This fact is known as Moore's Law \cite{Moore1965law}. 
Most transistors on a chip are used 
to store information in the cache memories. Since the time required\index{time!to access memory} 
to access cache's\index{cache} information is 
much lower than that of other memories (see fig. \ref{fig:memories}), this integration of transistors
makes computations faster. Processors released in the years 2008-2011 have typically of the order
of $10^9$ transistors, and their surface is on the order of squared centimeters. The integration of circuits
can increase the performance of all simulations, since the lower the time to access the information
to deal with, the lower the total required execution time.
\item  {\bf Clock-rate increase}: Until recently, the clock 
frequency\index{clock!frequency}\index{clock!rate} has followed its own
'Moore's Law', growing exponentially, although at a lower rate than the number of transistors per chip
(increasing about 1.75 times every two years). 
This growth has recently  (c. 2006)
collapsed \cite{expendingmooresdividend} (see fig. \ref{fig:integration_evolution}) because the higher the
clock-rate\index{clock!frequency}\index{clock!rate}, the larger the power\index{power consumption}
consumption, which scales with the cube of the clock-rate\index{clock!frequency}\index{clock!rate} 
\cite{future_of_microprocessors,multicore_math}, and higher rates would require 
cumbersome cooling systems. The increase of clock-rate has a direct effect on a computer's performance,
since in principle every operation takes a given number of clock cycles, and reducing the time of
every cycle will reduce by the same factor the total execution time. 
As an example, in fig. \ref{fig:testsjar}.A) we present the relation with the inverse of the
clock-rate of the total time required to evaluate
a self consistent field (SCF) iteration in the calculation of
the ground state of a system of five atoms (CH$_4$). In these tests, the Density
Functional Theory (DFT) code Octopus \cite{Cas2006PSS,Mar2003CPC,octopuslessi}
was used in an Intel(R) Core(TM) i5 CPU 750 computer.

\item  {\bf Pipelining}:\index{pipelining} This consists of making different functional
parts of the CPU\index{CPU} perform
different stages of a more complex task. For example, assume an instruction consists of 5 stages, 
and each has to be performed by a different part of the CPU\index{CPU} after the previous stage has been completed.
If every stage takes one clock cycle, then to perform the whole instruction would require 5 cycles. 
But if every part is giving one result per clock cycle, up to 5 instructions could be run every 5 cycles
(the optimal performance for pipelining is one instruction per cycle). Splitting instructions into many stages
can therefore increase the speed of calculations. Pipelining enables 
greater clock-rates\index{clock!frequency}\index{clock!rate}, although, as stated above, clock-rate
is limited by the power consumption. Modern processor\index{processor}s are strongly pipelined, and 
some of them divide basic instructions into over 30 stages. 
\item  {\bf Superscalarity}:\index{superscalarity} This is the capability of CPUs\index{CPU}
to provide more than one result per clock
cycle. It is essentially based on hardware replication. A superscalar CPU is capable of finding and 
decoding several instructions per cycle (commonly 3 to 6 nowadays).
This can be done because its registers can take information
from several levels of the memory hierarchy at a given cycle. In addition, several ALUs do work simultaneously. 
Superscalarity is also partly based on pipelining.
The availability of very fast caches\index{cache}, which can perform over one load or store operation per cycle,
also improves superscalarity. The compiler should take advantage of the superscalarity features of a
computer. 

\item  {\bf Multicore\index{core} architecture for processors}: In order to overcome the
limitations given by the sequential 
nature of the Von Neumann model, a powerful solution is to include not only one, but several CPUs 
(core\index{core}s) per socket\index{socket}\footnote{A socket is the physical
package where the multiple core\index{core}s are joined.}, thus
forming multicore\index{core} processor\index{processor}s\index{multicore processor}.
This new paradigm uses the additional available transistors to put new computation units to work, rather
than to try to make a single core\index{core} faster \cite{expendingmooresdividend}.
Multicore\index{core} 
solutions are employed more and more in common PCs\index{PC}.
They are also a common solution to circumvent the problem of the high energy consumption 
in supercomputers\index{supercomputer} \cite{IBM_meteorology,multicore_math}, since multicore schemes enable
the performance of the processor\index{processor} to be increased even if
the clock-rate is lowered.
The multicore\index{core} processor\index{processor} architecture shares the increasing available
transistors (being doubled 
every two years according to Moore's Law\index{Moore's Law}) among the core\index{core}s. 
The inclusion of more core\index{core}s, however, also carries some inconveniences.
For example, the growth of the number of 
core\index{core}s per chip implies a reduction in both the main memory
bandwidth\index{bandwidth} and the cache size available for each core\index{core}.
Another inconvenience of multicore architecture is that
the presence of many CPUs\index{CPU} simultaneously solving the same problem implies that 
the code of the programs should be devised to provide them with the appropriate parallel instructions, which
they should execute at the same time. 
The recent trend to use multicore\index{core} processor\index{processor}s\index{multicore processor} can be
appreciated in fig. \ref{fig:integration_evolution}.
It is worth remarking that while pipelining and superscalarity are perfectly compatible with
the Von Neumann model, the multicore\index{core} architecture is not. It is a modification of that paradigm, 
which entails new rules for the flow of information and the way in which the computer acts on it.
The multiple core\index{core}s of a processor can either lie on the same chip or not, 
but they lie on the same socket\index{socket}. Some examples of multicore\index{core}
schemes are displayed in fig. \ref{fig:comp3}.
Typical PCs\index{PC} have a single socket, while servers commonly contain two to four sockets, all sharing
the same main memory. Big parallel computers (see sec. \ref{Clusters}) commonly contain many sockets.
In fig. \ref{fig:testsjar}.B) we present an example of how increasing the number of cores reduces the
total time for a given task. The task of this example is the calculation of a SCF iteration in the calculation of
the ground state of a system of 180 atoms using DFT. For this calculation, the Octopus 
\cite{Cas2006PSS,Mar2003CPC,octopuslessi} code was used in the Jugene
(IBM Blue Gene architecture) cluster.

\begin{figure}[!ht]
\includegraphics[width=5.9in]{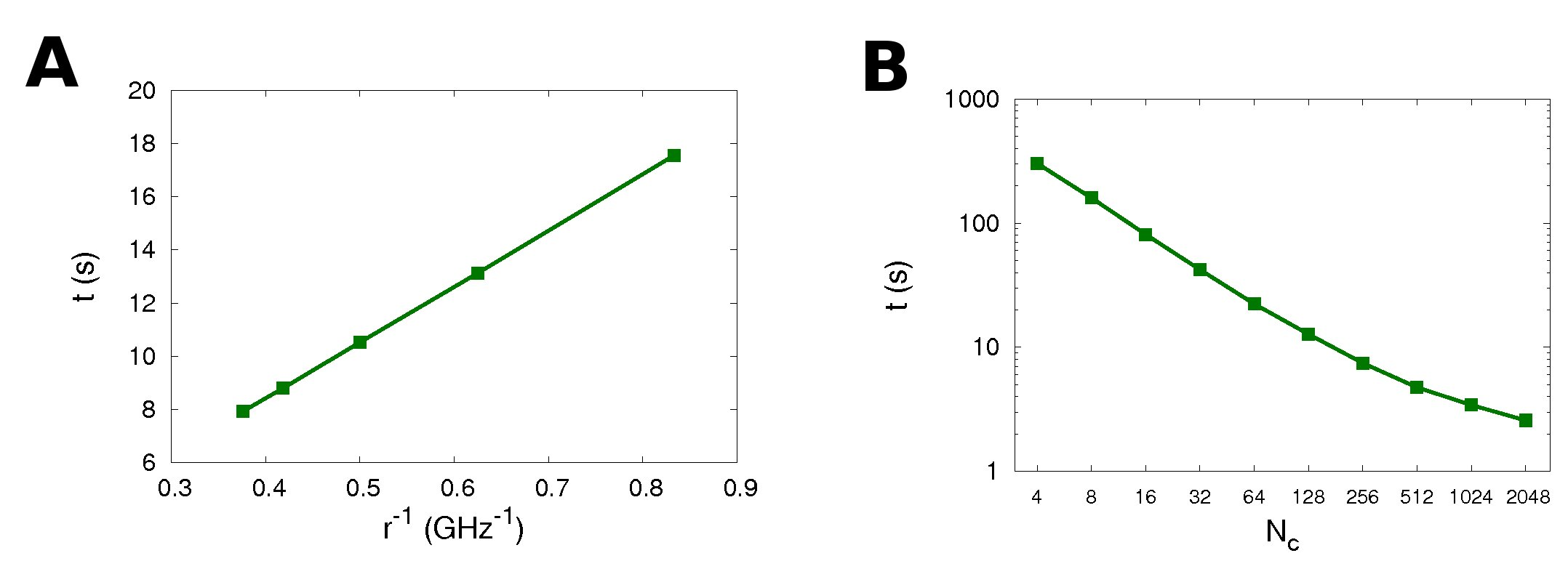}
\end{figure}

\begin{SCfigure}[][!ht]
\includegraphics[width=2.75in]{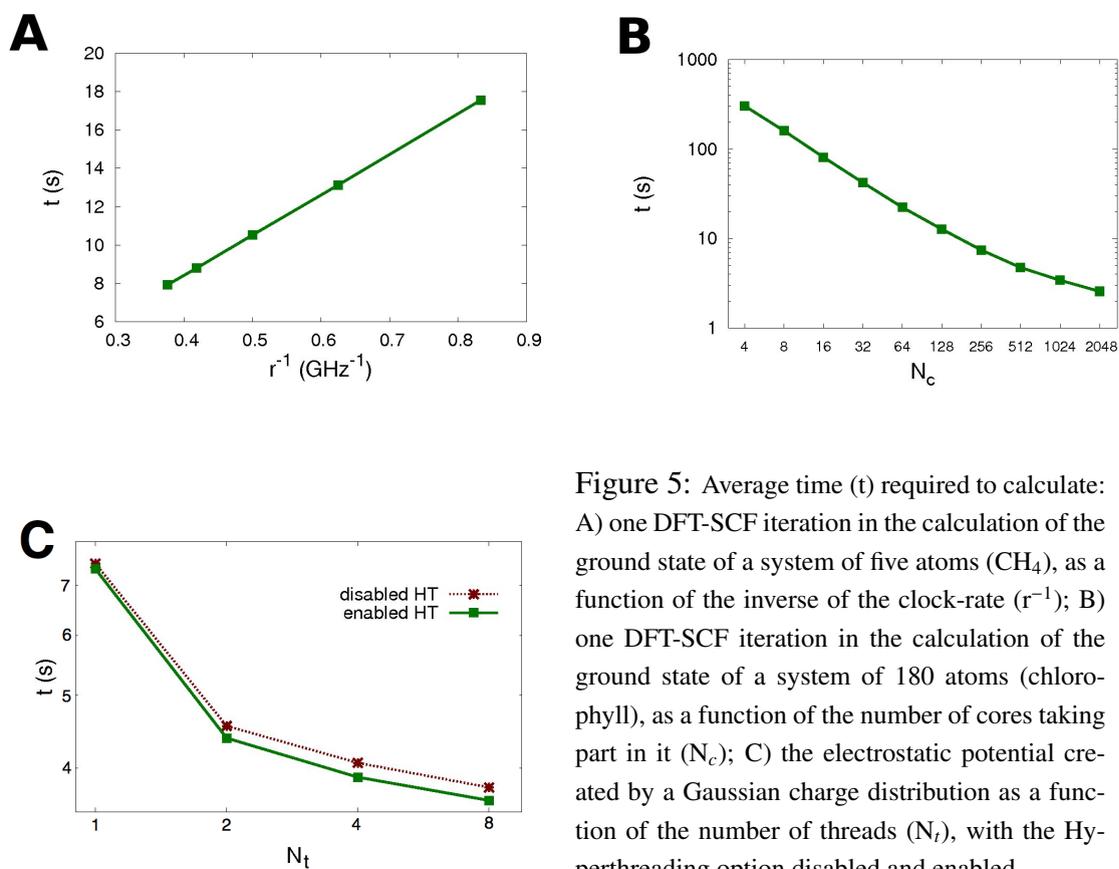}
\caption[short caption]{\label{fig:testsjar} {\footnotesize Average time (t) required to calculate:
A) one DFT-SCF iteration in the calculation of the ground state of
a system of five atoms (CH$_4$), as a function of the inverse of the clock-rate (r$^{-1}$);
B) one DFT-SCF iteration in the calculation of the ground state of 
 a system of 180 atoms (chlorophyll), as a function of the number of cores taking part in it (N$_c$);
C) the electrostatic potential created by 
a Gaussian charge distribution as a function of the number of threads
(N$_t$), with the Hyperthreading option disabled and enabled. 
}}
\end{SCfigure}

\item  {\bf Multithreading}: This is the ability of a core to execute instructions corresponding to 
several threads (several execution lines) which can be run in parallel (doing operations in different ALUs) or 
sequentially.
These threads can either be independent or mutually dependent. 
Multithreading is possible because a core has several sets of registers, each set storing information
of a given thread, although all threads in a core share common caches and depend on a single Control Unit. 
Multithreading is complementary to multicore architecture, since both parallelize the execution at 
distinct levels. In fig. \ref{fig:testsjar}.C) we show an example on the execution time of the influence of
using Hyperthreading (a concrete type of Multithreading). The example corresponds to the calculation of
the electrostatic potential created by a Gaussian charge distribution represented in a real space 
grid, and was run with the Octopus code in an Intel(R) Core(TM) i7-2600 CPU @ 3.40GHz machine. Note
that this example is just qualitative: the increase of performance of multithreading strongly depends
on the concrete problem and on how the software to tackle such problem is devised.

\item  {\bf SIMD instructions}:\index{SIMD instructions} SIMD (single instruction multiple data)
instructions enable data parallelism by performing arithmetic operations not on a single number, but on
a vector of them. The supercomputers\index{supercomputer} of the 1980s and early 1990s were based on this principle (although 
at a much larger scale) and they were called vector machines\index{vector machine}.
Although big vector machines\index{vector machine} are rare at present,
its principle is still useful for other computers (e.g., multimedia instructions are used for most general
purpose present-day processors). 
The use of SIMD instructions has proven to increase the performance of 
simulations in many fields such as, for example, molecular dynamics, fluid dynamics \cite{Latboltz} 
or astronomy \cite{simdastronomia}.
For example Machines using GPUs instead of
CPUs (thus executing the same instruction on many data simultaneously)
typically increase their performance in over one order of magnitude \cite{gpuhouston} in 
molecular dynamics simulations. Computers
not using GPUs can also take advantage of the SIMD paradigm, since more and more often they use 
registers of bigger size (say 128 bits vs 32 bits), which enables an ALU to do the same operation on 
various input variables simultaneously. The possibility to do this strongly depends on the software and 
the compiler, for the CPU must ''guess'' when (in what loop of the code) this can be done. 
An example of software developed to take advantage in the SIMD paradigm in molecular dynamics
calculations is NAMD, where increasing factors of 2 to 4 in the performance are common \cite{gromacs2001}.
Even for non-SIMD optimized programs, SIMD can result in an increase of performance. 
For example, the average value of the time required for a SCF iteration in the calculation of the
ground state of CH$_4$ (i.e., the example pointed when the \emph{clock-rate} was discussed) is about
a 5\% longer if SIMD are disabled.

\item  {\bf Out-of-order execution}:\index{Out of order execution} 
If the arguments of the instructions are not available in registers when they must be used (this can happen 
if the memory is too slow to keep up with processor speed) out-of-order 
execution can prevent computing units from being idle.
\item  {\bf Simplified instruction sets}:\index{Simplified instruction set} 
Paradigms of instructions sets which are rather simple
(as opposed to previous models) but can be executed much more quickly. The RISC (Reduced Instruction Set Computer)
paradigm was adopted during the eighties, resulting in efficiency increases.

\end{itemize}

All these improvements can be useful for simulation in essentially all the fields of
Physics, because they are quite general. The specific problem and the software to tackle it
will determine the point the increase in performance reaches.

The trends in the evolution of the degree of integration of processors\index{processor}, their clock-rates and
their numbers of cores are displayed in fig. \ref{fig:integration_evolution}.
We could say that the time of steady growth of \emph{single-processor\index{processor}}
performance seems to be over \cite{expendingmooresdividend}, which has spurred the semiconductor
industry to start a transition from sequential to parallel computers.
The introduction of multicore\index{core} processor\index{processor}s in 2004 (see purple line in
 fig. \ref{fig:integration_evolution}) marked the end of a
30-year period during which sequential computer performance increased from 40\% to 50\% yearly.
The trends displayed in fig. \ref{fig:integration_evolution} have lead to a reinterpretation of 
Moore's Law. In words of Jack Dongarra\index{Jack Dongarra} (coordinator of the 
top500\index{Top 500 list}
project): 
'The number of core\index{core}s per chip doubles every 2 years, while clock speed decreases (not increases). 
The number of threads of execution doubles every 2 years'.
This increase in the number of threads results from both the multi-core\index{core} solutions and from the 
hardware\index{hardware} modifications which enable CPUs\index{CPU} to work on other tasks when one executing thread is stalled
(for example, waiting for data) \cite{PARALLEL_HI-PERFORMANCE_COMPUT_CHEMISTRY}.
These capabilities increase computer performance, but require the software\index{software} to be well
suited to the hardware\index{hardware} to reach maximum performance. Although in the next years eventual 
physical limitations for technologies might make the thread increase trend stop, important 
advances are being achieved in the direction of core\index{core} integration. Intel has recently presented a 
prototype research chip\index{chip} which implements 80\index{multicore processor!80 core chip}
simple core\index{core}s, each containing two programmable floating-point engines.
This is the maximum single chip integration to date, and reaches TFLOPS\index{FLOPS}\index{FLOPS!TFLOPS} 
performance\footnote{http://software.intel.com/en-us/articles/developing-for-terascale-on-a-chip-first-article-in-the-series/?wapkw=\%28terascale\%29}
\cite{germann_Pflop_MD}.

\begin{figure}
\begin{center}
\includegraphics[width=5.6in]{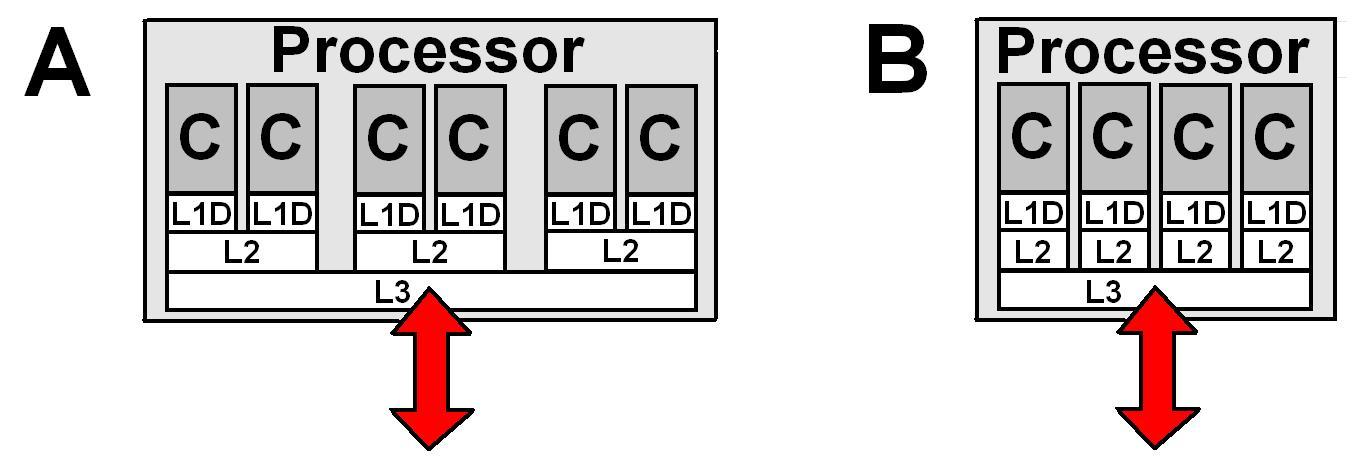}
\caption{\label{fig:comp3} {\footnotesize Two possible schemes for a multicore\index{multicore} processor\index{processor}\index{multicore processor}.
A) Hexa-core\index{core} processor\index{hexa-core}\index{multicore processor!hexa-core} 
chip with separate L1 caches\index{cache}, shared L2 caches for pairs of neighboring cores and a shared L3 cache for all core\index{core}s; 
B) Quad-core\index{core} processor chip\index{chip} with separate L1 and L2, but shared L3 cache\index{cache}
\cite{libro_HPC}.
}}
\end{center}
\end{figure}

\section{Parallel computers}\label{Clusters}
As stated above, the idea behind most powerful computers nowadays is the use of many core\index{core}s 
(which globally contain many CPUs\index{CPU}). A given program is divided into several execution lines (threads\index{parallel thread}) 
which are simultaneously run in the different core\index{core}s. Therefore, these cores solve the problem
in a cooperative way. The number of core\index{core}s 
used by a supercomputer\index{supercomputer} is continuously increasing in time. As stated 
above, the definition of 'supercomputer' or 
'high performance computer' is rather arbitrary\index{supercomputer!definition}, because these expressions refer to computers which
are much more powerful than common computers. Because of the technological advances, it is said
that today's supercomputers are tomorrow's PCs\index{PC} (see fig. \ref{fig:top500_evolution};\index{Top 500 list}
since a present-day laptop\index{laptop} is capable of doing about $10^{10}$ operations
per second (10 GFLOPS),
it is as powerful as a supercomputer was 14 years ago). The use of standard computer components to build
supercomputers (the so-called commodity clusters\index{cluster supercomputer}) has made these high performance
machines accessible for many research groups throughout the world.

\begin{figure}[!ht]
\begin{center}
\includegraphics[width=5.9in]{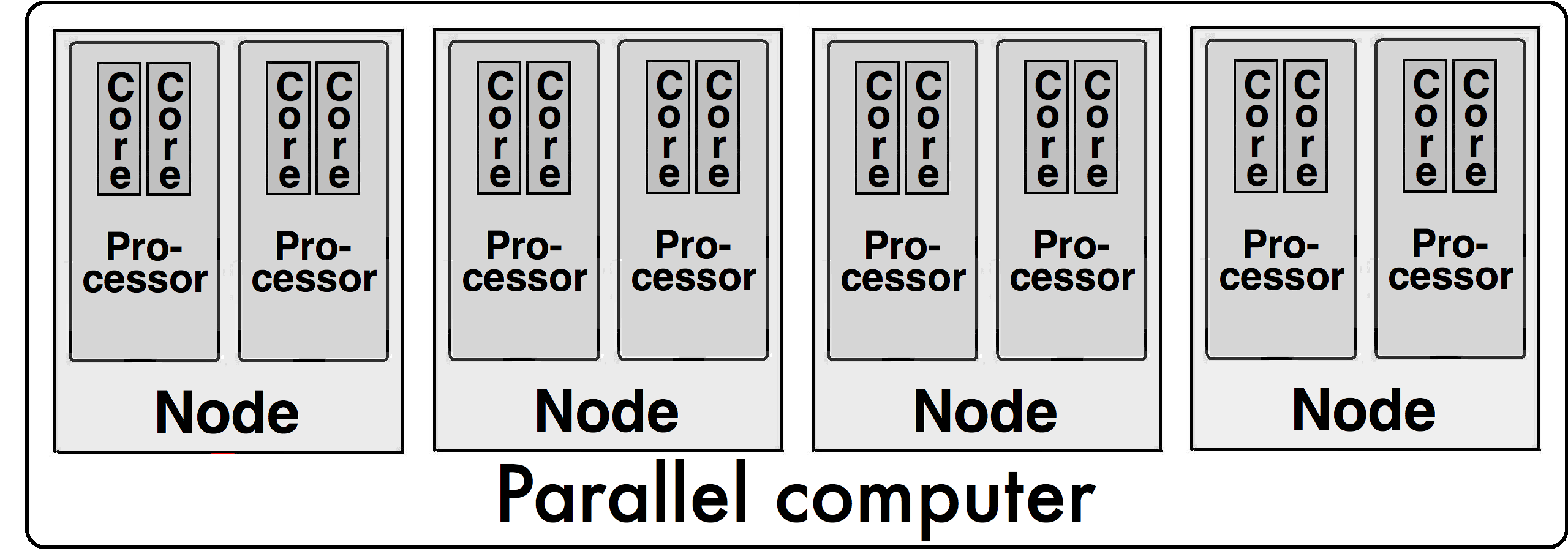}
\caption{\label{fig:comp12} {\footnotesize Scheme of a parallel\index{supercomputer} computer\index{parallel computer supercomputer!scheme}.
It can be divided into four 
levels (core\index{core}, processor\index{processor}, node\index{computing node} and parallel computer
itself). Each core\index{core} contains a CPU, the basic computing unit.
}}
\end{center}
\end{figure}

We call a parallel computer\index{parallel computer} a collection of connected 
core\index{core}s lying in different nodes, which are connected by networks.
The cores of an operating parallel computer work simultaneously in a cooperative manner. 
A parallel computer is formed following a scheme such as the one displayed in fig. \ref{fig:comp12}. It can be
decomposed into four levels: core\index{core}s, processor\index{processor}s, node\index{computing node}s
and the parallel computer itself. Components inside a given 
level have distinct features (e.g., bandwidth\index{bandwidth}, latency\index{latency}, etc.)
when communicating with other components
inside or outside its own module (e.g., a core\index{core} communicating with one 
core\index{core} in the same processor\index{processor}
will do it much more rapidly than with a core\index{core} lying in other processor\index{processor}).
We can summarize the four levels of a parallel computer as follows\footnote{The following definitions are commonly 
accepted, but not universal. In the field of computer architecture there is some inconsistency in the
nomenclature. In several contexts what we call a core is called a processor, and what we call merely 
a processor is called a multiprocessor chip or a multicore processor.}:
\begin{itemize}
\item Core\index{core}: It contains one Control Unit, and one or several 
arithmetic-logic units (and therefore one CPU). For example, cores devised under the Intel Core
microarchitecture have 
three ALUs \cite{intelarch2011}. A core can run one or several execution threads simultaneously.
\item Processor\index{processor}: Integrated circuit which contains one or several 
cores and is lying on one semiconductor layer (chip).
One or several processors are inserted in one socket\index{socket}.
\item Node\index{node}: Set of processors which share a common main memory, and commonly
also share other resources, such as a hard disk drive or a network connection.
It usually consists of one motherboard\index{motherboard} where 
there are several sockets\footnote{In some (rather rare) cases, the so-called twin motherboards, two
different sets of processors joined to two different main memories can lie on one motherboard.}.
Communication among processors in a given node is rather fast, and is provided by buses\index{bus}.
\item Parallel computer\index{parallel computer}: It comprises all nodes and the communications among them, which are provided by
networks such as Infiniband\index{Infiniband} or Gigabit Ethernet\index{Gigabit Ethernet}.
\end{itemize}

\begin{figure}[!h]
\begin{center}
\includegraphics[width=5.9in]{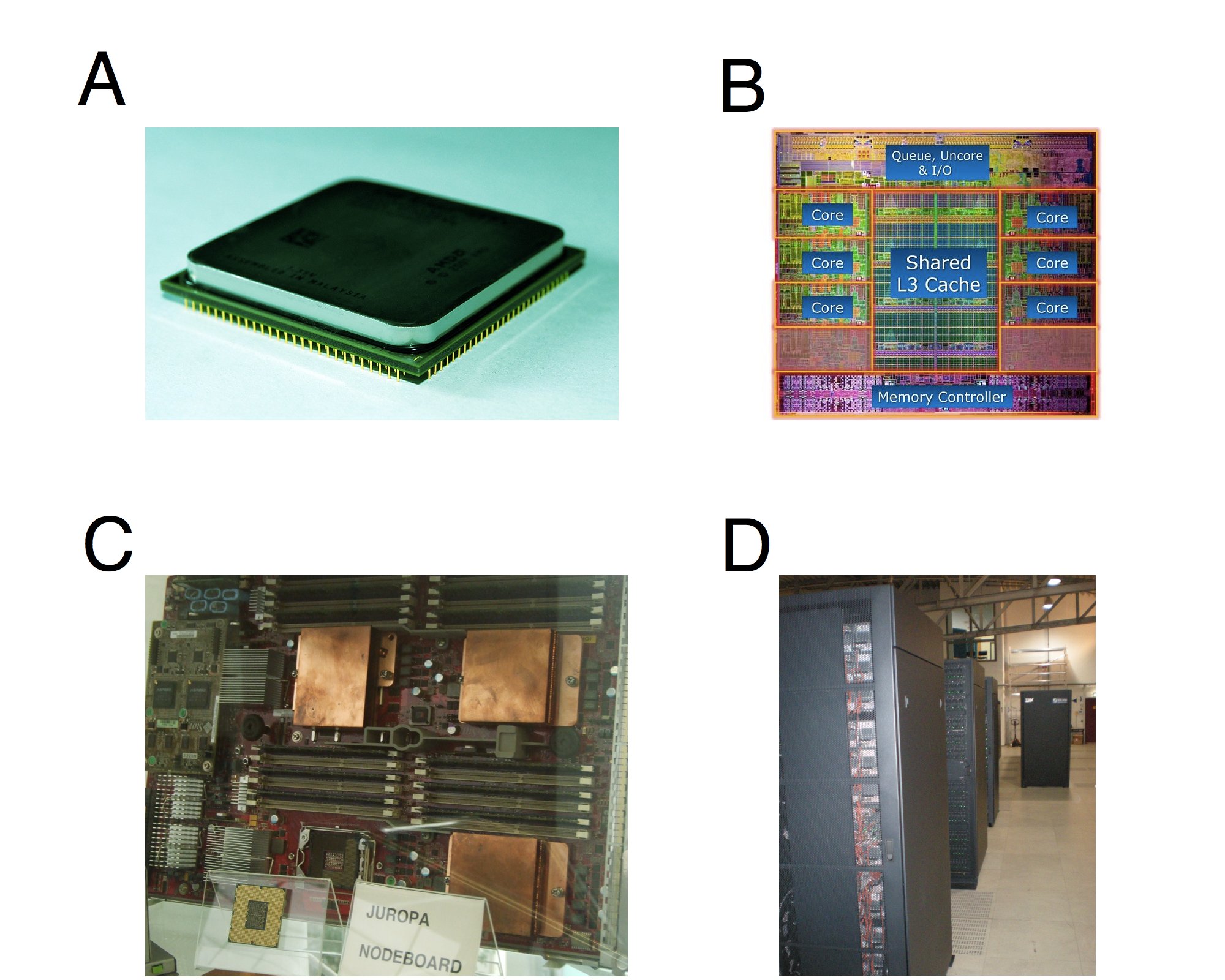}
\caption{\label{fig:levelscluster} {\footnotesize Examples of the constituents of a typical parallel computer. 
A) Processor (Dual-Core Athlon 64x2 - 6400plus);
B) Inner scheme of a processor 
(Intel Core i7-3960X Extreme Edition 6-core processor); C) Node (the image shows the motherboard of a node of the 
Juropa cluster at the JSC, lacking the hard disk);
D) Cluster (the image shows parallel computers at the JSC, J\"ulich, Germany). Image C) courtesy J. Alberdi.
        Image A) Copyright (c) 2011 Pablo Garc\'ia Risue\~no.
        Permission is granted to copy, distribute and/or modify this image
        under the terms of the GNU Free Documentation License, Version 1.2
        or any later version published by the Free Software Foundation;
        with no Invariant Sections, no Front-Cover Texts, and no Back-Cover Texts.
        A copy of the license can be found in \url{http://www.gnu.org/licenses/fdl-1.2.html}.
}}
\end{center}
\end{figure}

In fig. \ref{fig:levelscluster}, images corresponding to the different levels
enumerated above are displayed. A) displays
a processor, which contains two cores (each containing several control and arithmetic-logic units). 
B) displays the scheme of a 6-core processor. A motherboard (C),
together with devices joined to it, form a node. Several nodes linked by 
interconnection networks form a parallel computer (D).

Networks connecting the constituents of a parallel computer have a critical importance for
the parallel performance of applications in it. This is
because data transfer typically is the dominant performance-limiting 
factor in scientific code \cite{libro_HPC}.
The most important network characteristics that need to be taken into account in order 
to produce efficient parallel code are
its network topology (the way the node\index{computing node}s are connected) and its
network bandwidth\index{bandwidth} and latency\index{latency}
(see fig. \ref{fig:memories}). These features have an important influence on the performance of the parallel computer. 
Examples of common topologies\index{network!topology} are ring, grid, torus in 2 or 3
dimensions or tree \cite{PARALLEL_HI-PERFORMANCE_COMPUT_CHEMISTRY}.

We will now briefly describe various types of existing parallel supercomputers\index{supercomputer}. 
We will first
introduce the essentials of parallel computers, 
and then some fundamentals of other devices, like GPUs, special-purpose
computers and heterogeneous computers,
and of other types of computation, like the cloud computing and grid computing.

Parallel computers are commonly classified into two paradigms: shared-memory and distributed 
memory\index{distributed-memory machine}\index{shared-memory machine}. 
The latter are also known as \emph{clusters}\index{cluster}.
Both operate under the MIMD paradigm, i.e., multiple instructions are given to 
multiple core\index{core}s, which deal with multiple input
variables (data). Other systems\footnote{Please note that the classifications presented in this
paper not universal.}, like GPUs, vector machines\index{vector machine} 
or some microprocessor\index{processor}s
follow the SIMD paradigm (see sec. \ref{hardwarebasics}), and they execute a single instruction on 
a set of multiple data.

\begin{figure}[!h]
\begin{center}
\includegraphics[width=2.2in]{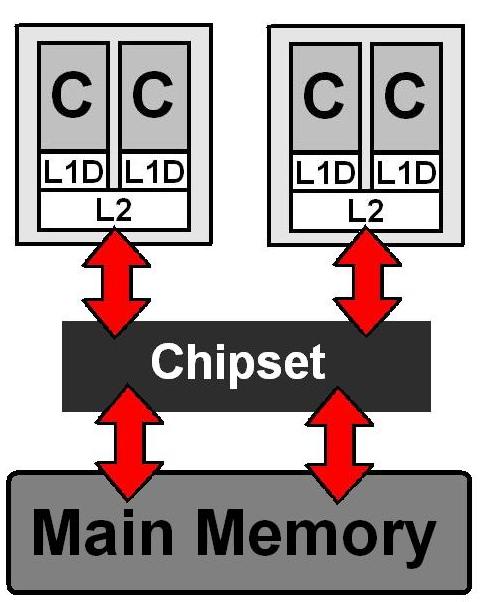}
\caption{\label{fig:comp4} {\footnotesize Scheme of a shared-memory\index{shared-memory machine} UMA machine consisting of two
dual-core\index{core} processor\index{processor}s \cite{libro_HPC}\index{dual core}\index{multicore processor!dual core}.
}}
\end{center}
\end{figure}

A shared-memory parallel computer is defined \cite{libro_HPC} as a system in which a 
number of CPUs operate in a common shared physical address space.
Important versions of this paradigm are:
\begin{itemize}
\item  {\bf UMA}\index{UMA shared-memory machine}\index{shared-memory machine!UMA} 
(uniform memory access): The access time of all processor\index{processor}s 
for the main memory is (essentially) the same. This is attained with communication devices having
the same latency\index{latency} and bandwidth\index{bandwidth}.
\item {\bf ccNUMA} (cache\index{cache}-coherent Non-uniform Memory Access): 
The main memory\index{main memory}\index{memory!main memory} is 
physically distributed across the various processor\index{processor}s, but the circuits (logics) of the machine 
make this set of main memories to appear as only one large memory, so the access to different parts 
is done using global memory addresses. The access time is different for different processor\index{processor}s and
different parts of the memory, as in a distributed-memory computer.
\end{itemize}
The scheme of the UMA paradigm can be found in fig. \ref{fig:comp4}. In UMA architectures, a device called
\emph{chipset} controls the information flow between the main memory\index{main memory}\index{memory!main memory}  and the core\index{core}s.
The simplest example of UMA is the dual core\index{core} machinery\index{dual core}\index{multicore processor!dual core}
that recently has become very popular \cite{germann_Pflop_MD}. 
The complexity of the circuits required to keep the access time uniform, at present, limits
the largest UMA systems with scalable bandwidth\index{bandwidth} (the NEC SX-9 vector node\index{computing node}s) to
sixteen sockets\index{socket} \cite{libro_HPC}. 

\begin{figure}
\begin{center}
\includegraphics[width=4.0in]{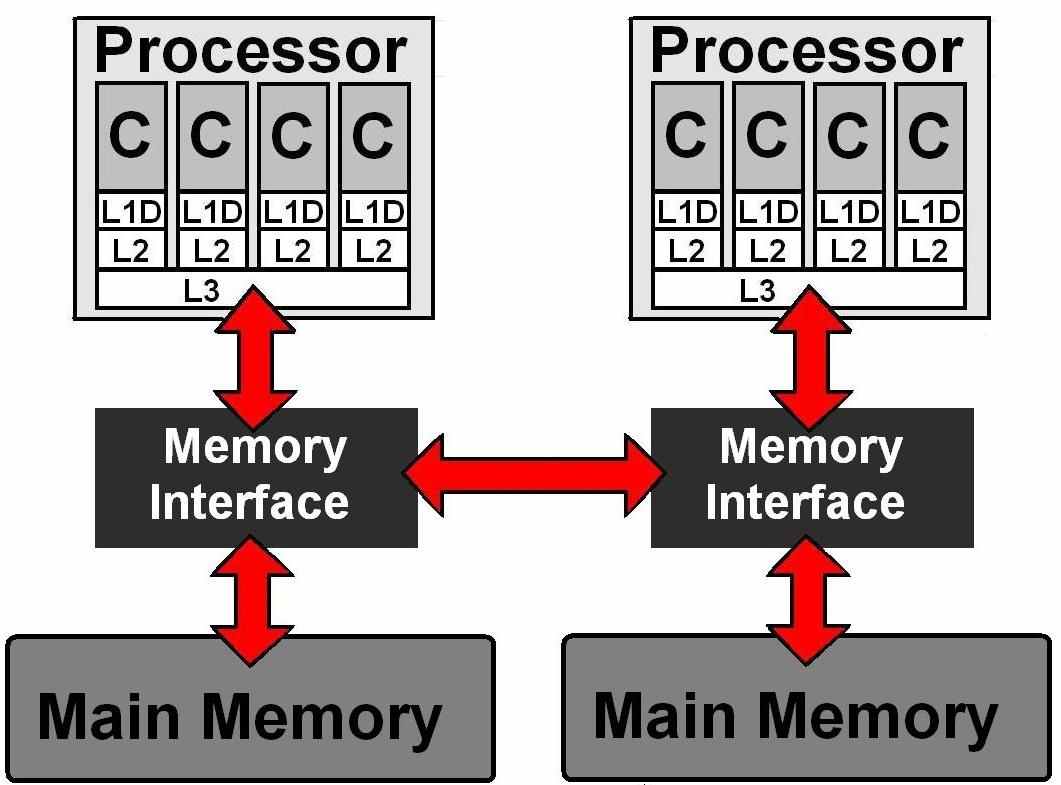}
\caption{\label{fig:comp6} {\footnotesize Pattern of a ccNUMA system with two locality domains 
and eight core\index{core}s. Red arrows indicate information flow.
The information is managed by a memory interface. Memory interfaces are 
connected via coherent links (for example, the central red arrow in this picture) \cite{libro_HPC}.
}}
\end{center}
\end{figure}

\begin{figure}
\begin{center}
\includegraphics[width=6.1in]{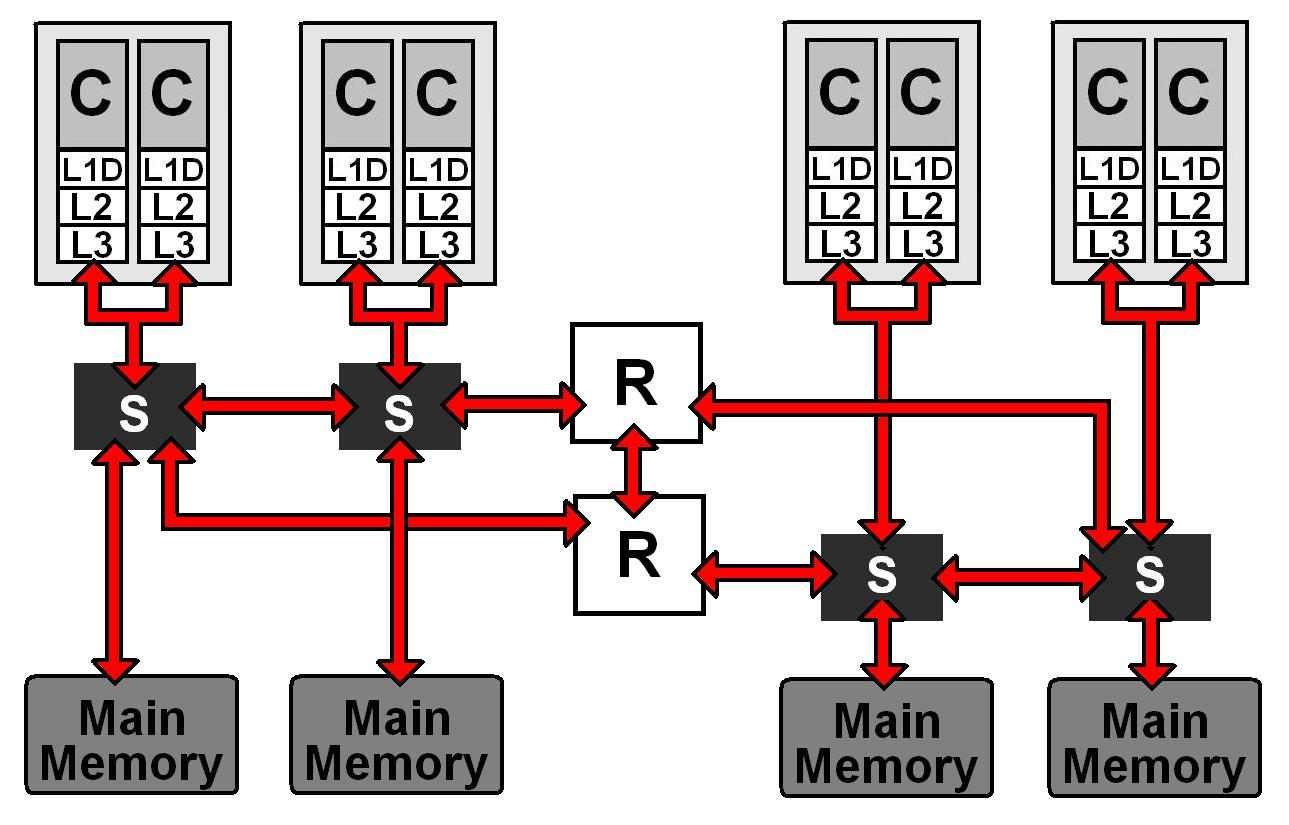}
\caption{\label{fig:comp7} {\footnotesize  A ccNUMA system with four locality domains, each comprising one 
socket\index{socket} with two core\index{core}s. Each socket is linked to a communication interface S,
and the traffic of information is managed by some routers R \cite{libro_HPC}.
}}
\end{center}
\end{figure}

Typical patterns of shared-memory ccNUMA\index{shared-memory machine!ccNUMA} machines
are displayed in figures \ref{fig:comp6} and 
\ref{fig:comp7}. A ccNUMA computer consists of several local domains, whose memories are locally 
connected (each local domain being basically a UMA). 
The memories of different local domains communicate via so called coherent links. This architecture
is appropriate for large shared-memory machines, but is more often used to build small 2- or 4-socket
node\index{computing node}s in supercomputers.

An architecture which is more widely used to build large supercomputers\index{supercomputer} (up to thousands of core\index{core}s) is the
one whose scheme appears in fig. \ref{fig:comp7}. 
Each processor\index{processor} socket\index{socket} is connected to a communication interface (S), which
provides memory
access to the proprietary NUMALink network\index{network!NUMALink}. The NUMALink network uses routers (R) for
connections with nonlocal units. The asymmetry of this design makes the access times\index{time!of memory access} very variable
for different processor\index{processor}s and different memory positions. ccNUMA machines also have the drawback that 
two or more different processor\index{processor}s may try to access the information at a given memory position simultaneously,
and thus they would compete for its resources. In addition, the input/output human interface is connected with only one local
domain.

Pure distributed-memory\index{distributed-memory machine} computer schemes would include one 
main memory\index{main memory}\index{memory!main memory}  per core\index{core}, each being connected
to all or part of the others. Such patterns, however, are seldom found because of their
price/performance features. Most of the so-called
distributed-memory machines (which are indeed a large portion of supercomputers\index{supercomputer})
are actually hybrid models, i.e.,
distributed-memory machines whose building blocks 
(node\index{computing node}s) are shared-memory-like devices \cite{PARALLEL_HI-PERFORMANCE_COMPUT_CHEMISTRY}. 
\begin{figure}
\begin{center}
\includegraphics[width=6.1in]{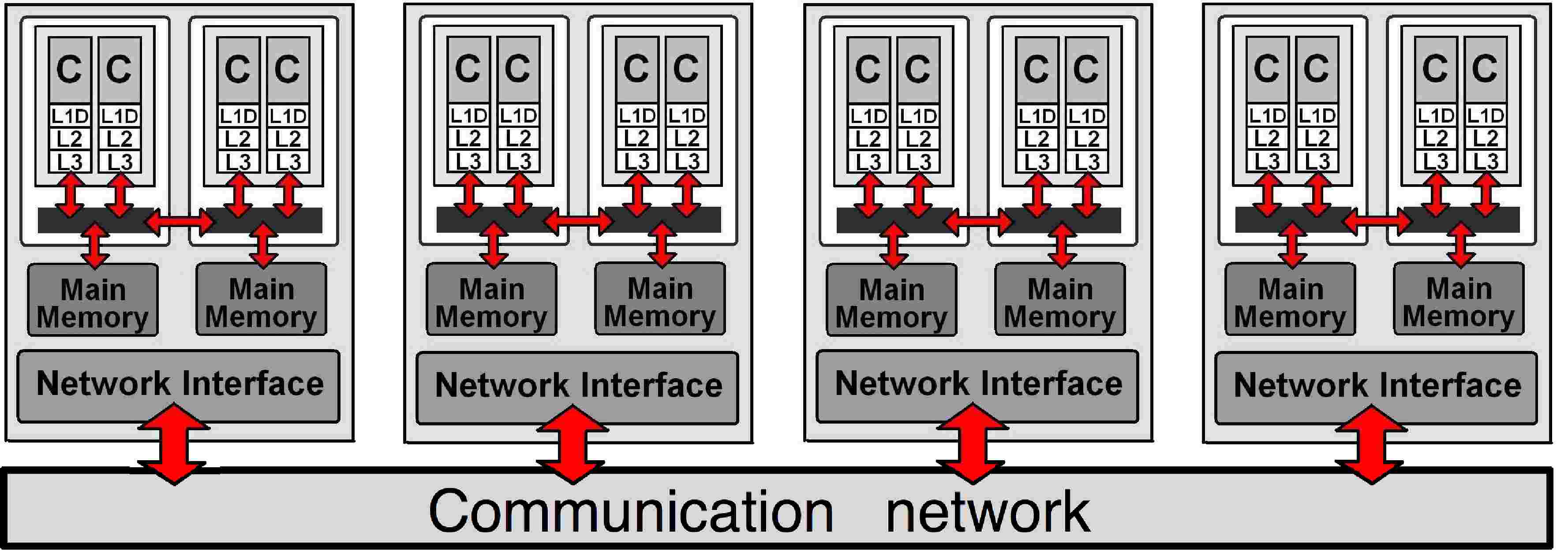}
\caption{\label{fig:comp10} {\footnotesize Pattern of a distributed-memory\index{distributed-memory machine} computer.
Red lines indicate information flow \cite{libro_HPC}.
}}
\end{center}
\end{figure}
When the code being executed at a given core\index{core} requires information which is stored in the memory 
belonging to another node\index{computing node} or in another disk memory location,
the core\index{core} must send a request for that information.
The request reaches the core\index{core}
after travelling through the network\index{network} connecting the parts of the cluster\index{cluster supercomputer}.

For supercomputation
purposes, these connections were in the past made with the Gigabit Ethernet technology (see fig. \ref{fig:memories}).
At present Infiniband\footnote{\url{http://www.infinibandta.org}}\index{Infiniband}\index{network!Infiniband} 
has become very popular (see fig. \ref{fig:top500}).
It is worth stressing that at present there exist serious inconveniences with information transfer
devices,
because they transfer information much slower than standard CPUs\index{CPU} perform operations. For example, 
Infiniband connections have a latency\index{latency} of the order of microseconds ($10^{-6}$ s). Since present-day 
clock-rates\index{clock!frequency}\index{clock!rate} are of the order
of GHz, and a floating-point operation\index{floating-point operation!number of clock cycles}
takes of the order of tens of clock cycles, a typical operation
can take about $10^{-9} \sim 10^{-8}$ s. Therefore, transferring the information of the result of 
an operation between nodes will be at least 100 times slower than performing the operation. 
In addition to latency, the bandwidth of the network can also be a limiting factor. 
Delays produced by nonzero latency and finite bandwidth are the reasons why
minimal feedback among different node\index{computing node}s is pursued by algorithm developers \cite{foldingathomeresults}.
Feedback among core\index{core}s with fast connections, such as those belonging to the same processor\index{processor}s, 
need not usually be avoided \cite{future_of_microprocessors}.
It is said that, qualitatively, the speed of a serial computer is determined by the caches, while
the speed of a parallel computer is especially determined by the speed of communications among nodes
\cite{HenYYYYBookappI}. Standard values of latencies and bandwidths of caches and networks can be 
viewed in fig. \ref{fig:memories}. There exist some ways to compensate for the big difference
of latency between caches and networks, like the \emph{communication latency hiding}, 
that consists of overlapping communication with computation (or with other communication) 
\cite{HenYYYYBookappI}. 

\begin{figure}[!ht]
\begin{center}
\includegraphics[width=3.8in]{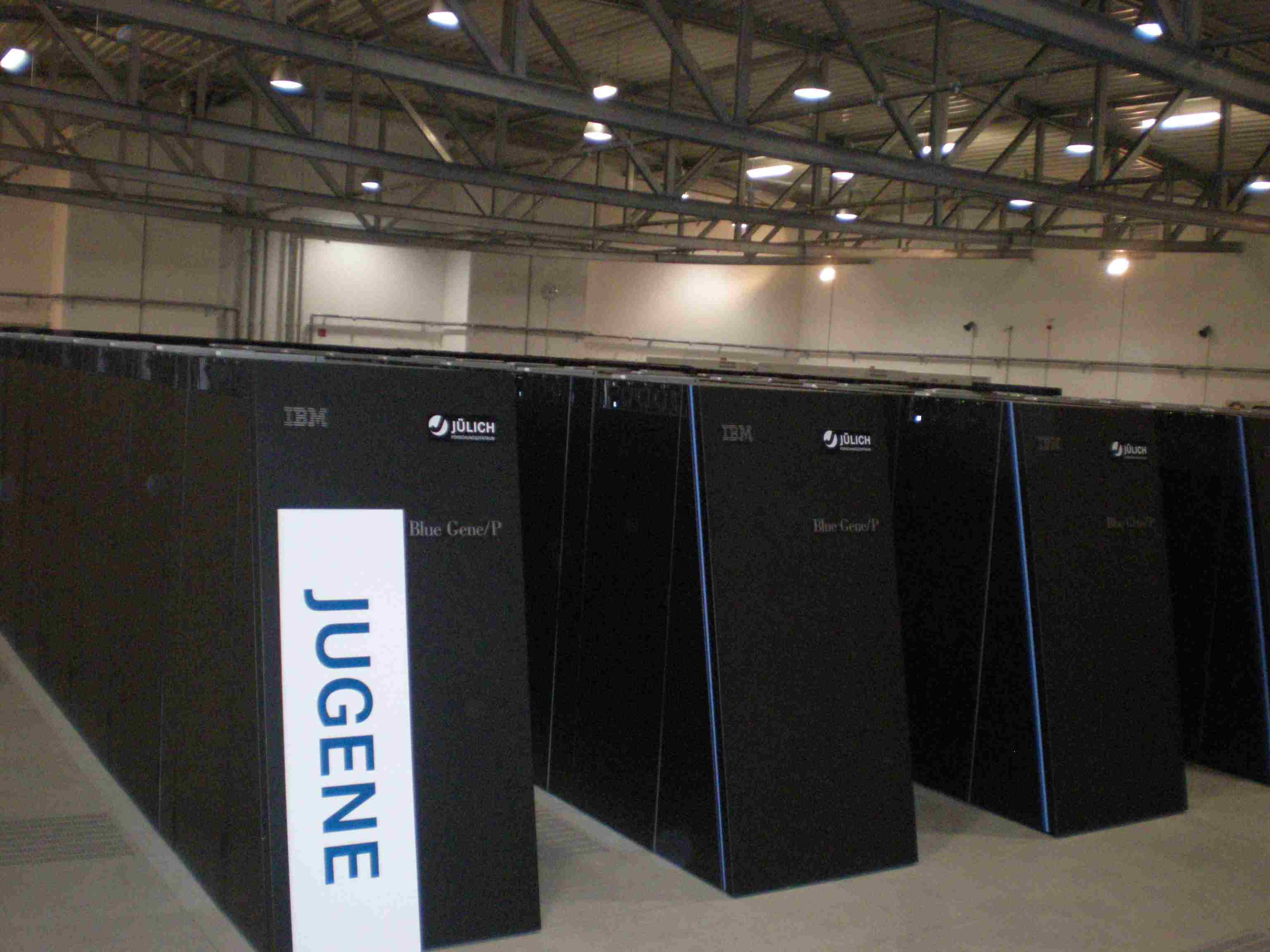}
\caption{\label{fig:Jugene} {\footnotesize
The cluster supercomputer Jugene\index{Jugene}\index{supercomputer!IBM Blue Gene}\index{supercomputer!Jugene} (IBM Blue Gene / P technology), in the JSC (J\"ulich, Germany) in 
Sept. 2010. In that moment, it consisted
of 294,912 core\index{core}s, what made it the most powerful computer in Europe, and the fifth most powerful
computer in the world.\index{IBM Blue Gene}
}}
\end{center}
\end{figure}

The dominant HPC architectures\index{dominant HPC architectures} at
present and for the foreseeable future are comprised of node\index{computing node}s which
are (shared-memory) NUMA machines themselves, 
and which are connected with the rest of the node\index{computing node}s following a 
distributed-memory\index{distributed-memory machine} pattern \cite{MPI_OpenMP_HPC}.
A popular architecture of distributed-memory computers is IBM's Blue Gene (see fig. \ref{fig:Jugene}). 
In Blue Gene\index{IBM Blue Gene}\index{supercomputer!IBM Blue Gene} \cite{bluegenebook}, processor\index{processor}s
lie in the vertices of a cubic grid. Each processor\index{processor} communicates with its six nearest neighbours
in a 3D torus topology (the last processor\index{processor}, in the border of the
grid, is linked with the first one, in all three directions). 

If a problem is to be solved in a cooperative manner by the various cores of a cluster, 
the code to be executed must be written so that the workload is shared among them.
The MPI (Message Passing Interface) protocol \cite{guialargampi} is appropriate for writing codes that run in parallel in 
distributed-memory machines, as well as
for shared-memory\index{shared-memory machine} machines. 
The increasing popularity of MPI at present comes from its simplicity and its availability of standard libraries.
The parallelizations of many Physics and Chemistry simulation programs are based on MPI, although other 
parallelization paradigms are also used \cite{PARALLEL_HI-PERFORMANCE_COMPUT_CHEMISTRY}.
Despite its remarkable advantages, the MPI paradigm has an important drawback. Information transfer
between two parallel threads\index{parallel thread} requires both of them to reach a synchronization\index{synchronization},
i.e., one must execute
the instruction for sending information (MPI\_SEND)\index{MPI\_SEND} and the other one must execute the instruction for
receiving it (MPI\_RECV)\index{MPI\_RECV}.
In this way, the sender core\index{core} will be idle until its information is be received.
The limitations of the MPI scheme can be overcome by providing facilities for a process
to access data of another process without that process' direct participation \cite{PARALLEL_HI-PERFORMANCE_COMPUT_CHEMISTRY}.
Performance in shared-memory\index{shared-memory machine} machines can be increased by using the OpenMP
interface. Since most supercomputers\index{supercomputer!hybrid} 
have hybrid shared-distributed-memory\index{distributed-memory machine} architectures, 
mixed use of different programming codes, including both MPI and OpenMP\index{OpenMP} is advisable \cite{MPI_OpenMP_HPC}.
However, many programmers use only MPI\index{MPI}, for the sake of
simplicity in their codes.

\section{Hybrid and heterogeneous models}\label{hahm}

A supercomputer can either be formed by the repetition of processors\index{processor} of the same kind,
or by different 
ones, such as general purpose processors, graphics processing units or special-purpose chips, 
among others. In the former case, the architecture is called \emph{homogeneous}, while in the
latter case, it is called 
\emph{heterogeneous} \cite{hetero_chips}\index{heterogeneous computer}.
Heterogeneous machines can have advantages with respect to homogeneous 
machines in power consumption\index{electric power}\index{power consumption}, 
efficiency of data transfer and parallel speedup.
The Cray supercomputers\index{Cray supercomputers}\index{supercomputer!Cray}\footnote{\url{http://www.cray.com}}
are examples of heterogeneous machines, which
get high performance by using different kinds of processor\index{processor}s within the same cluster computer.

Using graphics processing units (GPUs \cite{hi-performance_comp_gpu_nvidia}) as a part of (heterogeneous) 
clusters\index{cluster supercomputer!heterogeneous} is a more and more popular solution. 
GPUs\index{Graphics Processing Unit}\index{GPU} were originally devised to perform fast 
calculations on data for creating images. However, 
in recent times, they have become celebrated in the context of supercomputing. 
This is because they are capable of treating large amounts of data (vectors of data, rather than 
single variables) simultaneously. 
A GPU can perform some given tasks several orders of magnitude faster than a CPU.
For example, in some problems of data analysis,
GPUs performance can gain about a factor of 200 with respect to CPUs\index{CPU}
performance\footnote{\url{http://www.hpcwire.com/hpcwire/2011-03-29/comparing\_gpus\_and\_cpus.html}}.
For general-purpose computation issues, GPUs can either be included in 
motherboards\index{motherboard} together with
CPUs, or
different types of devices (GPGPUs\footnote{General Purpose Graphic Processing Units.}
\cite{IBM_meteorology})\index{GPGPU} can be produced that can work without a CPU.
Full clusters can be built on GPGPUs instead of CPUs\index{CPU}.
As a sign of the current success of GPU-based solutions, it can be noted that four
out of the ten most powerful supercomputers (in the list of top500 as of June 2011\index{Top 500 list}) include 
GPUs.

A different kind of supercomputing facility is the one formed by the special-purpose 
computers\index{special purpose computer}.
These machines cannot deal with a very broad range of different tasks, as the general-purpose 
computers (PCs\index{PC}, laptops\index{laptop}, most supercomputers\index{supercomputer}, etc.) can. 
Instead, they are devised to execute
a limited set of algorithms, but with great efficiency. 
An example of a special-purpose 
computer is Anton\footnote{\url{www.deshawresearch.com}}, which is dedicated to protein 
molecular dynamics\index{molecular dynamics}.
Anton\index{Anton supercomputer}\index{supercomputer!Anton} was able to run a simulation\index{simulation}
corresponding to 1 ms (in a huge number of steps, since $\Delta t$ is of the order of fs in proteins), 
realizing the hard task of predicting the folding of a protein\index{protein}\index{protein!folding prediction}
\cite{Dro2010JCB,Sha2009XXX}. 
Other examples of special-purpose computers are Janus\index{Janus supercomputer}\index{supercomputer!Janus} \cite{special_purpose_janus},
which is used for simulation of spin\index{spin!glass} glasses, 
and GRAPE-6\index{supercomputer!GRAPE-6} \cite{special-purpose_astronomy}, for astronomy\index{astronomy}.

\begin{figure}[!ht]
\begin{center}
\includegraphics[width=5.8in]{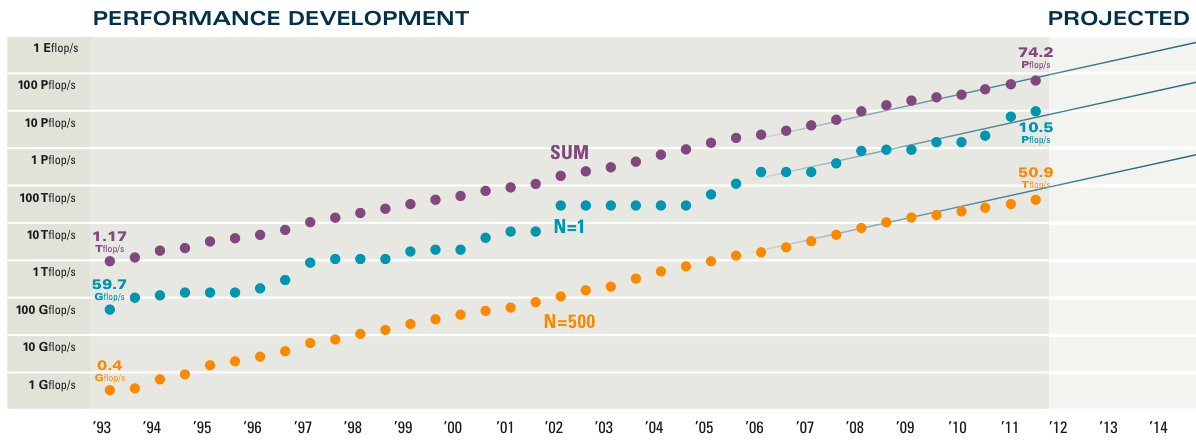}
\caption{\label{fig:top500_evolution} {\footnotesize\index{Top 500 list}
Evolution of computational capabilities: purple line: Aggregated power of all 500
most powerful computers in the world, according to the top500 list;
blue line: most powerful computer in the world,\index{most powerful computer in the world} 
according to the top500 list; orange line: 500th most powerful computer in the world.
Notice that the vertical axis is logarithmic. Source: top500 (\url{www.top500.org}).\index{Top 500 list}
}}
\end{center}
\end{figure}

\begin{figure}[!ht]
\begin{center}
\includegraphics[width=5.7in]{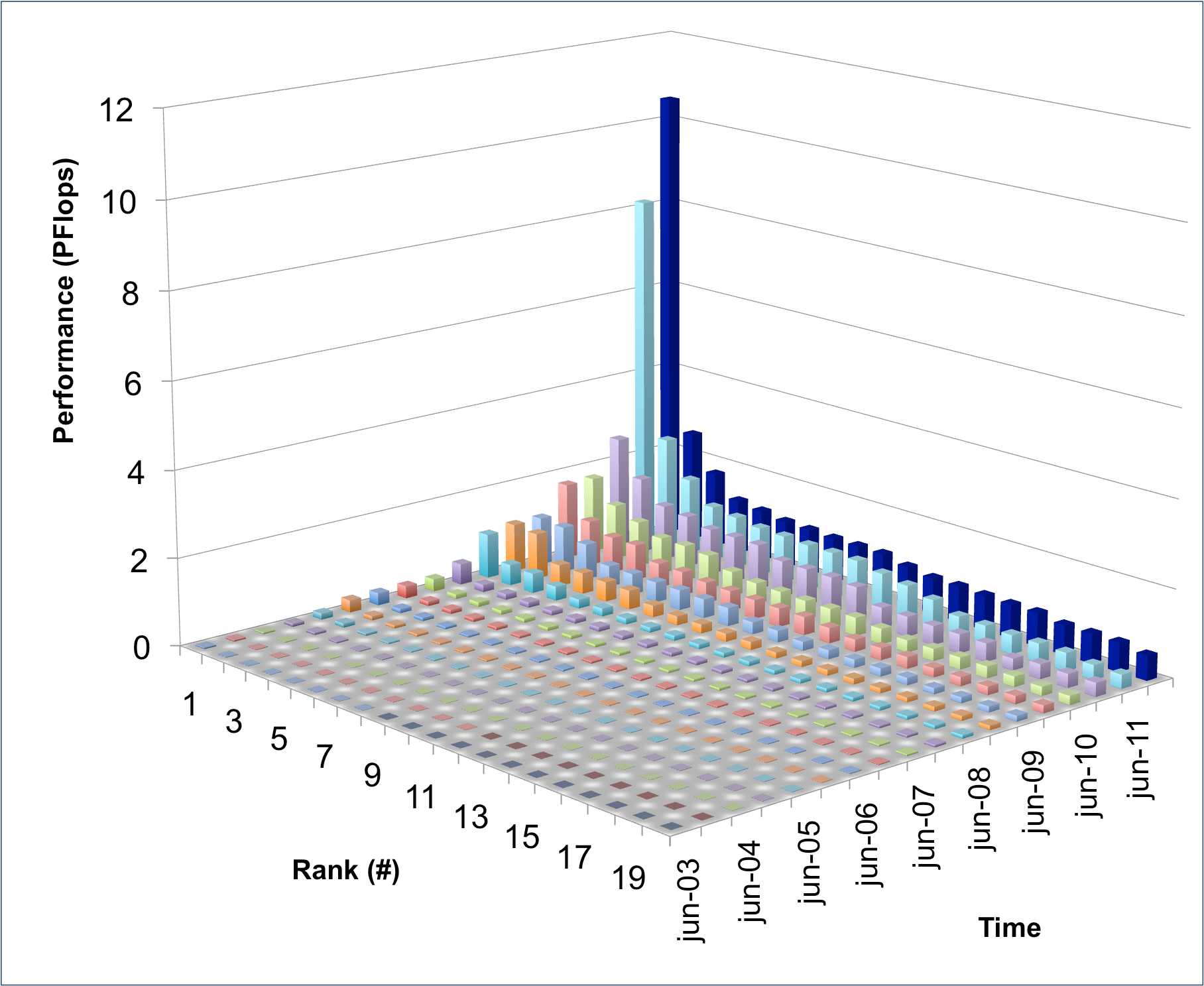}
\caption{\label{fig:top20_evolution} {\footnotesize
Evolution of the\index{Top 500 list} performance of the twenty most powerful computers on 
the world from June 2003 to November 2011. (Source: top500 ---$R_{max}$---). 
}}
\end{center}
\end{figure}

\begin{figure}[!ht]
\begin{center}
\includegraphics[width=6.0in]{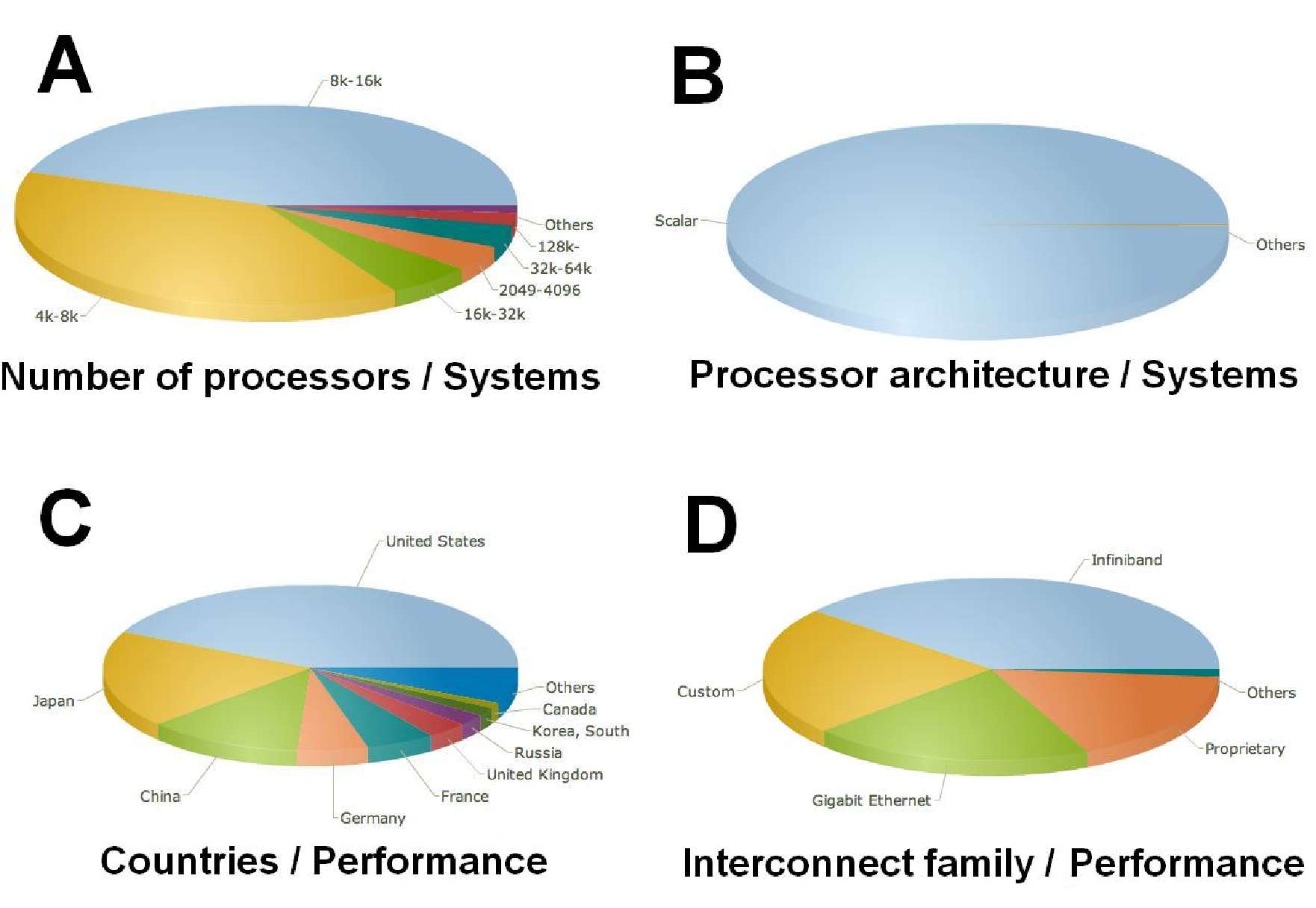}
\caption{\label{fig:top500} {\footnotesize
Some features of the most powerful computers in the world, according to the top500 list of June 2011.
The data labeled as 'systems' (upper row) means that every computer on the list enters the charts with 
weight 1. The data labeled as 'performance' (lower row) means that the weight depends on the 
computer's performance. A: Number of processor\index{processor}s; B: Processor architecture; C: Location of 
machines; D: Family of interconnect devices.
(Source: \url{www.top500.org}).
}}
\end{center}
\end{figure}

Fig. \ref{fig:integration_evolution} and some of the subjects
discussed so far can give us an idea of the evolution of 
computational capabilities of modern computers, which grow very rapidly, in an exponential way.
In order to quantify this evolution, twice a year a list of the 500 most powerful computers on the 
world, the top500 list\index{Top 500 list}, is released. In order to measure the performance of computers, a benchmark
(the HPLINPACK benchmark\index{HPLINPACK benchmark} \cite{linpackb2003}) is run.
The essential task of the HPLINPACK benchmarks\index{HPLINPACK benchmarks} is to solve a linear system
of equations $Ax=b$ whose coordinate matrix
$A$ is dense (i.e., contains no null entries) with partial pivoting.
Performance is measured in number of 
floating-point operations per second (FLOPS\index{FLOPS}) that the machine is able to perform.
It is worth remarking that special-purpose computers can hardly appear on the top500 list, because
it is essential to run the HPLINPACK benchmark\index{HPLINPACK benchmark},
and many special-purpose 
computers\index{special purpose computer} are not capable of doing it.
In fig. \ref{fig:top500_evolution} we can appreciate the exponential increase of supercomputing 
performance of the machines in the top500 list, both for individual supercomputers (the most powerful
one, and the one in the
500 position, appearing in the figure) and for the aggregated power of all computers on the list.
From 1993 to 2011, 
the combined computing power of the collection of all 500 most powerful computers in the world followed its 
own 'Moore's Law', since it increased about 1.9 times per year (almost twice as fast as the increase in 
integration predicted by Moore's Law).
By comparing charts \ref{fig:integration_evolution} and \ref{fig:top500_evolution}, we can 
appreciate that the trends in chip integration, clock-rate and number of cores per processor
and the power of high-performance computers are closely related.
In fig. \ref{fig:top20_evolution}, we can see
the evolution of the 20 most powerful computers during the last decade. The vertical axis of this
graph is not logarithmic, so the dramatic increase can be noticed in a clearer manner. 
In fig. \ref{fig:top500}, some features of the top500\index{Top 500 list} computers are summarized. 
Chart A shows
the number of core\index{core}s high-performance computers consist of. Chart B quantifies 
something we mentioned
earlier in this section, that vector computers\index{vector machine} are no longer popular,
and most modern computers are 
scalar (i.e., every CPU\index{CPU} deals with a scalar variable at a time, not with a vector containing many
values simultaneously). Chart C gives an idea of the computer power distributed by countries. Finally, 
chart D gives a notion on the network\index{network} communication devices which are used by most powerful 
supercomputers\index{supercomputer}.

The enhanced capabilities of current supercomputers\index{supercomputer} to perform operations very 
rapidly is very
useful for scientific calculations. Most modern algorithms and codes are based on repetition to take 
advantage of these capabilities. However, developing software\index{software developer} for massively parallel machines 
is harder than for the traditional sequential computers. This is because 
instructions should be provided to many units working simultaneously, and efficient information 
feedback among them needs to be implemented.
The growing complexity of computers, where many (not necessarily equal) processing units
are linked among them and also linked to a hierarchy of memories\index{hierarchy of memories}, 
each having a different access time, 
makes parallel programming a rather complicated task. The complexity of computers' architecture
is an extra obstacle for the programmers \cite{future_of_microprocessors}. More complex internal behaviour of 
core\index{core}s, including 
effects of pipelining and superscalarity makes code execution harder to comprehend in depth and 
their logic hinders parallelization, because human programmers find it hard to devise codes which
are well-suited for these architectures.
For example, the programmer can think that the code he writes is executed sequentially and in order in a core\index{core}, but this is 
no longer true (as stated when the Out-of-order execution was explained). Programmers are helped by 
modern compilers, which include several levels of optimization, which can do more efficient mapping
in data transfer and result in large efficiency improvements.
In order to devise more efficient software\index{software} codes, tracing and profiling software tools can be very 
useful \cite{profiling_mohr}.
Moore's Law is said to have enabled cheaper (more efficient) execution of programs, but at a higher
cost of developing new programs, by reason of the increasing complexity of computers \cite{expendingmooresdividend}.
In words of J. Dongarra\index{Jack Dongarra}, 'For the last decade or more, the research investment strategy has been 
overwhelmingly biased in favor of hardware\index{hardware}. This strategy needs to be rebalanced, since barriers to progress 
are increasingly on the software\index{software} side. Moreover, the return on investment is more favorable to software.
Hardware\index{hardware}\index{hardware!half-life} has a half-life measured in years, while software\index{software!half-life} has a half-life measured in decades. 
Unfortunately, we don't have a Moore's Law\index{Moore's Law} for software, algorithms and applications'. This does not 
mean that the aim of efficient algorithms has not been successful, but means that
modern algorithms and programs should be well-suited to current machine component features and computer
architectures (an example of this adaptation can be found in \cite{germann_Pflop_MD}, where the 
efficiency is increased by means of optimizations of the transfer of data).
Indeed, smart software parallelization schemes, including new paradigms, can be the only tool to improve
supercomputers' performance\index{supercomputer} 
when hardware capabilities no longer increase
\cite{future_of_microprocessors}.

A few years ago, machines supporting High Performance Computing have had to face an important obstacle:
the significant rise of 
power consumption\index{electric power}\index{power consumption}. The increase of clock-rates has 
boosted power dissipation \cite{libro_HPC}, because the power 
consumed by a core\index{core} scales with the cube of its clock-rate \cite{multicore_math}. This 
is attributable to the fact that
the power is proportional to the square of the voltage multiplied by the frequency, but the voltage itself is 
proportional to the clock frequency\index{clock!frequency}\index{clock!rate}. Power dissipation 
generates heat, which is an important drawback
because semiconductor\index{semiconductor} materials in computers require rather low temperatures in order
to work properly. To this end, cooling systems\index{cooling system} must be implemented. Computers in the 1980s did not need
heat sinks; the heat sinks used in the nineties were of moderate size; today's cooling systems\index{cooling system} 
are very big. In PCs and laptops\index{laptop} air cooling systems (fans) are usually enough, but in supercomputers\index{supercomputer}
water-based cooling systems may be mandatory in a nearby future \cite{future_of_microprocessors}.
The work of cooling systems entails serious power 
consumption\index{electric power}\index{power consumption} which must be added to that resulting
from the processor\index{processor}s' work.
As a result of these high power needs, the budget for electrical energy of the computing facilities of a research
group may easily be exceeded. 
As an example \cite{energy_saving_2}, a 16-core\index{core} processor\index{processor} with every core\index{core} consuming an average of 20 
Watts will lead to 320 Watts total power when all core\index{core}s are active, which will have a non-negligible
economic cost. Another example is given by Amazon.com \cite{Hamilton_Amazon}.
The energy bill of its data centers costs about the 42\% of the total budget of the center (the cooling\index{cooling system}
system consuming more energy than the operation of CPUs\index{CPU}).
Many ways to reduce the impact of excessive power consumption\index{electric power}\index{power consumption}
are being proposed \cite{energy_consumption,energy_saving_cloud}.
Architectures for many core\index{core} machines should be carefully devised to minimize it. 
A possible solution is to simplify processor\index{processor} designs. Using larger 
caches is another option, although there is a limit beyond which a larger cache\index{cache} will not pay off any
more in terms of performance, due to bandwidth\index{bandwidth} limitations. 
Multicore\index{core}\index{multicore processor} processor\index{processor}s are the solution 
which most present computer manufacturers prefer \cite{libro_HPC}.

\section{Distributed computing}\label{ditribcomp}
Data on the top500\index{Top 500 list} refers to well localized large supercomputers (like the one displayed in fig.
\ref{fig:Jugene}). 
However, in recent times, other solutions for HPC, such as \index{grid computing} grid computing, 
\index{cloud computing} cloud computing and volunteer computing, have become very popular. 
These three more modern ways of computing are said to be ways of
\emph{distributed computing}\index{distributed computing}\index{computing!distributed}.  
Distributed computing is based on the concept of 
doing high performance calculations using geographically distant machines, which has been 
enabled with the advent of the internet and high-speed networks.
Computers participating in a given problem can lie thousands of kilometers
away from each other, but they can share information through the internet. 

grid computing \cite{libro_grid_computing_ch1} uses
geographically distant node\index{computing node}s
to solve a given problem simultaneously in a cooperative way. 
grid computing capabilities
are usually managed by a given organization, and the computational resources (mediated by physical machinery)
which support the calculations 
are provided by different supporting institutions and organizations, which can be
companies, research groups, laboratories, universities, etc. 
A management committee distributes the computational capabilities at every moment among the requests of
different groups of users\index{user!of grid computing}.
The groups of people taking part in grid computing projects for solving problems are usually
called \emph{virtual organizations} since these groups are frequently heterogeneous, being 
formed by many people from different organizations which are geographically distributed. 
The essential aspect of a virtual organization is that it is formed for a specific project.
Virtual organizations can act either as producers or consumers of resources (or both). Various
virtual organizations involved in a grid computing project are mutually accountable; i.e., 
if one misbehaves, the others can cease sharing resources with it.
Since many computer core\index{core}s
throughout the world are working together, much computing power can be 
accumulated, which enables solving many problems whose solution may not be feasible even in the 
most powerful supercomputing clusters\index{cluster supercomputer}.
This generates vast amounts of data, which spurs the creation of large collective databases \cite{data_management_2005}.
Large grid computer facilities are often used by a large number of users, which
helps to match the demand of the computational resources with their availability.

It is also worth noticing that grid computing projects commonly operate under open-source\index{software} 
software 
standards, which eases the development of software applications and the cooperation among groups.
A popular software\index{software!Globus Toolkit} package to manage grids is the Globus Toolkit, 
including the GRAM software
as a tool for the users. Grid facilities, as well as cluster computers, frequently run in Linux 
operating systems.

\begin{figure}[!ht]
\begin{center}
\includegraphics[width=3.8in]{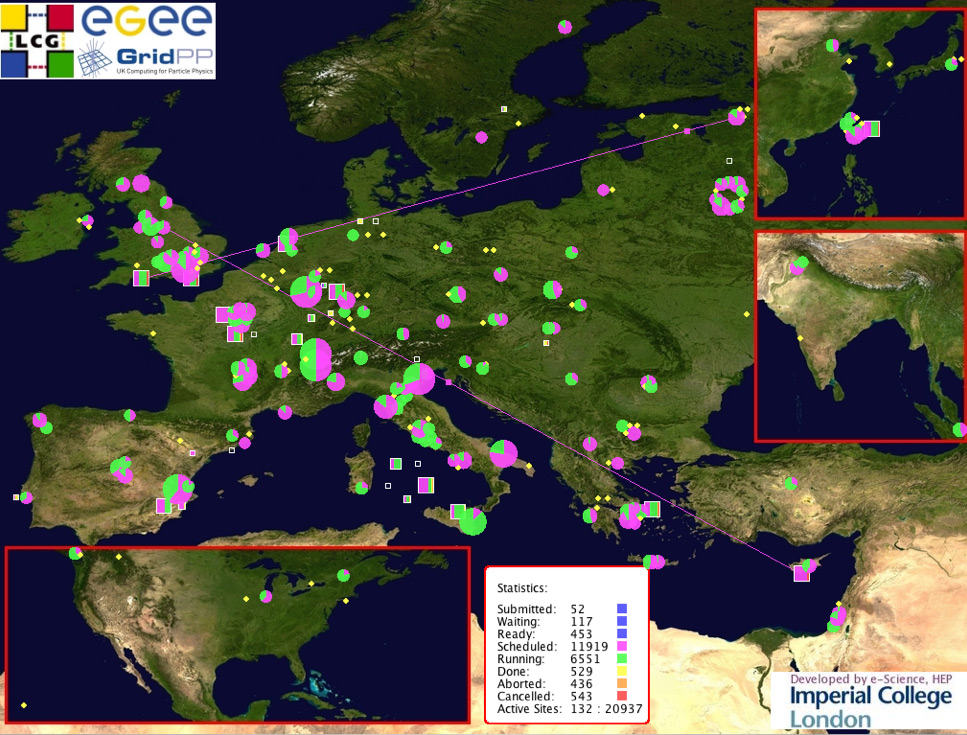}
\caption{\label{fig:grid} {\footnotesize
A scheme of geographical distribution of computing node\index{computing node}s of the egee grid (\url{http://www.eu-egee.org}). 
Current grid computing projects often involve facilities in many countries.
}}
\end{center}
\end{figure}

grid computing\index{grid computing} has been successful in numerous research fields, such as drug design,
biomolecular simulation,
engineering and computation for industry, Chemistry, 
Geology (e.g. earthquake simulations) or meteorology \cite{libro_grid_computing_ch1,gridDNA}. It also plays
an important role in
Particle Physics. For example, the Large Hadron Collider of the CERN sends huge amounts of experimental 
data to its associated
grid facility (the Worldwide LHC Computing Grid\index{Worldwide LHC Computing Grid}, WLCG),
so that many scientific groups throughout the world can analyse the data.
WLCG\footnote{\url{http://lcg.web.cern.ch/lcg}} involves over 140 computing centres in 35 countries, and 
includes several national and international grid projects.

Recently, another kind of distributed computing, volunteer computing, \index{high performance computing!volunteer computing} \index{volunteer computing} 
\index{volunteer computing} has become a useful tool for scientific purposes. volunteer computing
consists of using the computation power of machines which were neither devised nor purchased to 
do scientific calculations, but for use in daily life. Common PCs\index{PC} and laptops\index{laptop}
connected to the internet,
like those in millions of homes, can perform calculations to solve scientific problems. 
It is only necessary that the owner of the computer agrees and installs the appropriate software\index{software} 
(this is the reason why this kind of computing is called volunteer computing).
As stated in \cite{foldingathomeresults},
there are hundreds of millions of idle PCs potentially available for use every moment, and
the majority of them are strongly underused. Moreover, while the complexity and the network efficiency have
increased following their own Moore's Laws\index{Moore's Law}, the number of computer users\index{user!computer} has increased at even a higher rate
during the last decades, which makes the potential capabilities of volunteer computing huge \cite{foldingathomeresults}.
volunteer computing has produced many remarkable scientific results in the last decade \cite{grid_kondo}.
Some examples of volunteer computing are the Ibercivis project\footnote{\url{http://www.ibercivis.net}},
seti@home\footnote{\url{http://setiathome.berkeley.edu}}
---for the search of extraterrestrial intelligent life---
and folding@home\footnote{\url{http://folding.stanford.edu}}\cite{foldingathomepaper}
---for statistical calculations of molecular dynamics\index{molecular dynamics} trajectories 
for models of biological systems---. The last one is a particularly good example of how important scientific 
results can be produced with volunteer computing for problems which are unaffordable for other
HPC schemes \cite{foldingathomeresults}.
%
%
Volunteer computing projects often rely on the BOINC\index{software}\index{software!BOINC} open-source
software\footnote{\url{http://boinc.berkeley.edu}},
which is also appropriate for grid computing.
Although grid computing and volunteer computing share some features, there is a key difference between them.
grid computing is commonly \emph{symmetric}
while volunteer computing is commonly \emph{asymmetric}. This is, in the former, one organization 
can borrow resources one day, 
and supply them the next; in the latter, contributors (particular computer owners) commonly
just provide resources to the project.

Distributed computing has become a powerful research tool because of its numerous advantages.
Nevertheless, it also presents drawbacks.
The main one is that the distant geographical distribution of the different computing node\index{computing node}s
makes feedback among them much slower than if they were located in the same building. This fact makes
distributed computing not so useful for problems which require
frequent information feedback among computing units. Such a disadvantage has a physical limitation
which cannot be circumvented. For example, let us assume that
two computing node\index{computing node}s\index{node!|see{\emph{computing node}}} lie 3,000 km away from each other. Having
perfect communication, with data moving at the speed of light, information would take about 10 ms 
to travel between them. 
If the clock-rate of the processors involved in the grid is of the order of GHz, and
an operation requires, for example, 10 clock cycles, then the operation will require
about $10^{-8} \sim 10^{-9}$ s to be performed. This means that 
over one million sequential operations
could be performed by one computing node\index{computing node} before it could receive more data from the other one to 
continue with its calculations.
Other drawbacks of grid and volunteer computing are
security ones, since secure data transfer is much harder to maintain in complex connections 
via the internet. In addition, since results come from various distinct resources, 
they may require frequent overhead to check their validity.

Apart from grid computing and volunteer computing, cloud computing
\cite{cloud_computing_astronomy,cloud_foster,grid_kondo} also supplies computational capabilities for 
scientific calculations by connecting to remote machines via the internet. This is performed by 
powerful computers that companies dedicate for this purpose (usually for a fee). 
In cloud computing, a set of virtual servers work together to satisfy user requests, enabling
interactive feedback and taking advantage of the available computing capabilities to maximize their
use. Cloud computing has some advantages with respect to other ways of computing. For example, it
enables the user immediate access to computational resources without the need 
to obtain approval from an allocations committee
and the service can be provided without human interaction
with the service provider. Cloud computing enables the use of software without the need for purchasing
a licence or installing it, and the user does not need to have strong qualifications in 
software or infrastructure management. Cloud computing can be classified into three models \cite{cloud_foster}:
\begin{itemize}
\item Software as a service (SaaS): The user can run the available software, but he cannot 
install new programs or configure the operating system.
\item Platform as a service (PaaS): The user can install new programs, but he cannot 
act on the operating system.
\item Infrastructure as a service (IaaS): The user is enabled to configure the infrastructure; 
he can install new software, configure the operating system, the network, etc.
\end{itemize}

At present, several companies offer cloud computing resources at competitive prices. 
Downloading vast amounts of data generated from calculations done in the cloud, however, is 
customarily relatively expensive. 

\index{computation|)}

\section{Intrinsic limitations to accuracy and efficiency}\label{workcompsci}

When performing scientific calculations,
both software developers and software users should try to avoid some important issues related to methodology,
which are commonly related to accuracy\index{accuracy} and execution time\index{execution time}.
We can call \emph{accuracy} the similarity between the result of a given calculation and the
hypothetical result that would be obtained if we were able to perform the same calculation without
any numerical error.
When calculating physical or chemical
quantities, the accuracy\footnote{The word 
\emph{precision} is sometimes used as a synonym
of \emph{accuracy}, but we prefer not to use it because, more properly, 'precision' means low standard
deviation from a mean value, which need not be the actual sought value.} is essential, because
a lack of accuracy makes results unreliable. 
The accuracy can be lowered by many sources of error that exist for calculations performed in computers.
For example,
real numbers are usually represented in floating-point notation\index{floating-point notation}, 
each number being encoded in a finite number of bytes\index{byte}, usually 4 or 8. 
A real number is said to
be represented in single precision\index{single precision} if 4 bytes (i.e., 32 bits, 32 
binary figures) are used, and it is said to be
represented in double precision\index{double precision} if 8 bytes are used.
In the popular IEEE arithmetic, the 32 bits of a single-precision number are distributed as follows: 
1 bit for the sign, 8 bits for the exponent,
and 23 bits for the fraction \cite{HenYYYYBook}. 
The number $e$ stands for the $e$xponent, while $f$ stands for the \emph{fraction}\index{fraction}. Together,
they represent a real number whose absolute value is $1.f \times 2^e$ (the notation $1.f$ means that
$f$ is the fraction to be added to 1; for example, if the fraction is 0.25, then the number to multiply
by $2^e$ ---the significand\index{significand}--- is 1.25).
A simple way to represent integer numbers in binary code is to use the first digit for the sign
(0 for -, 1 for +) and the $n$-th digit to be multiplying $2^{n-1}$. In this code, for example, the
8-digits binary number 01010011 = - 
($1 \times 2^6 + 0 \times 2^5 + 1 \times 2^4 + 0 \times 2^3 + 0 \times 2^2 + 1 \times  2^1 + 1 \times 2^0$) = 
-83.
However, the integer number for the exponent ($e$) is sometimes represented in the biased
exponent, which is different than the one just presented. 
In the biased exponent, the number that $e$
represents is the number its figures represent (according to the encoding just presented)
minus a given number, which commonly equals $2^{bit \ number -1}$.
For example, in this notation $0011 \neq 3$, $0011 := 3 + (-2^{4-1}) = -5 $. 
In the encoding used for the fraction part $f$, each figure
is to be multiplied by the inverse power of 2 corresponding to its place. For example, 
$f=0100 \ \rightarrow \ f=  0\times 2^{-1} + 1\times 2^{-2} + 0\times 2^{-3}+ 0\times 2^{-4} = 0.25$.

The finite size of variables stored by computers implies that a finite number of binary figures 
is used to represent every number. This implies that
not all the existing real numbers, but only a subset of them, 
can be exactly represented 
in a computer. In most cases, when representing one number in a computer we are using
not that very number, but the closest number to it that the computer can represent (in a given notation). 
Every rational number with a denominator which has a prime factor which is not a power of 2 has
an infinite binary expansion.
Let us see an example.
If we are dealing with single precision, floating-point, real numbers 
according to IEEE arithmetic, then the number 0.1, which is exact in the decimal notation,
is periodic in binary notation. Its value will be 
\begin{equation}
e=01111100=-4; \quad  f=1001100110011001100110011001100 \ldots \nonumber
\end{equation}
We can notice that the represented single precision binary number is not exactly the number
we wanted to represent:
\begin{eqnarray}
& 1 & \ \ 01111100 \quad 100 1100 1100 1100 1100 1100 \ = \nonumber \\
& + & 2^{-4} (1 \ + \ 1\times 2^{-1} + 1\times 2^{-4} +1\times 2^{-5} +1\times 2^{-8} +1\times 2^{-9} +1\times 2^{-12} + 1\times 2^{-13} + \ldots)  \nonumber \\
& = & 0.09999999404 \neq 0.1 \ . 
\end{eqnarray}
Errors made in this way are called \emph{machine precision errors}\index{machine precision error}.
If we do an operation with several numbers, each represented in limited precision, then the errors
can accumulate. 
For example, consider we want to calculate the product of two (actual) numbers $a_{a}$ and 
$b_{a}$. We cannot represent them exactly in the computer, but we can only represent $a_{r}$, 
$b_{r}$ such that $a_{a}=a_{r}+r_a$ and $b_{a}=b_{r}+r_b$.
If we calculate their product in the computer, we will get $(a\times b)_{r}=
a_{r}\times b_{r} + e_{a\times b}$, being 
\begin{equation}
a_{a} \times b_{a} \simeq a_{r}\times b_{r} + 
a_{r}\times r_b + b_{r}\times r_a \ .
\end{equation}
If we multiply many numbers, the individual errors can accumulate. This makes that any other operation
or sequence of operations using real numbers can propagate errors as well. In summary, results in a 
computer calculation are usually not exact, but they depend on the precision\footnote{Do not confuse this 
(computer precision) with statistical precision, related to typical deviations.}
(number of bytes used to represent a real number)
chosen.

The order in which the operations are performed in an algorithm makes the machine precision errors
propagate in different ways.
There exist virtually an infinite number of algorithms capable of doing some given calculation, for
there exist many mathematically equivalent ways to do the calculations aimed 
at reaching a desired result. When these algorithms are implemented 
to do calculations in a computer, the way in which the errors accumulate can be quite different.
For example, 
if $a$ and $b$ are real numbers, then for a computer $a/b \neq a\times (1/b)$, by a slight margin.
We can appreciate it in the following example. Let aux1 be a double precision variable 
(i.e., it has about 16 decimal digits of precision),
the code\footnote{These results are for a specific computer and programming paradigm and will differ somewhat in 
other cases.}
\begin{algorithmic}
\STATE{ aux1=1234567891011121314150000000000.;}
\FOR {($i=0;i<220;i++$)}
 \STATE{ aux1=aux1/1.3456789333;}
\ENDFOR
\end{algorithmic}
gives a result 53.0145642653708151, while the code
\begin{algorithmic}
\STATE{ aux1=1234567891011121314150000000000.;}
\FOR {($i=0;i<220;i++$)}
 \STATE{ aux1=aux1*(1./1.3456789333);}
\ENDFOR
\end{algorithmic}
results in 53.0145642653714972. Hence, both results differ in the 14th figure.
This is merely a simple example, but it can be useful to notice that
every algorithm implementation has its own way to propagate errors.

In addition to the finite precision errors,
computers also introduce \emph{soft errors}\index{soft errors}, which are defined as
errors in processor execution that are due to electrical noise or external 
radiation rather than design or manufacturing defects \cite{softerrors}.

Apart from errors arising from technological limitations, the algorithm chosen to solve a 
given problem has also its own sources of error. Every algorithm
has its own mathematical definition and is based on a given level of theory.
For example, iterative algorithms require an starting 
guess, which may lead to wrong results if it is not appropriate, and they also require a criterion to decide when iterations
should stop. 
The set of equations used to tackle 
a system can also be an important source of error, since every system requires appropriate equations and
appropriate input parameters.

Apart from the accuracy, the other main limiting factor in computer simulation is the execution time.\index{execution time} 
Nowadays, we do know equations which describe small scale phenomena quite well, but their solution for
complex systems is cumbersome, and often unaffordable. The numerical complexity of the solution 
of simulation problems usually increases
with the size or complexity of the system tackled. This numerical complexity\index{numerical complexity} 
can be measured
with the number of required operations. Some examples of this can be
\begin{itemize}
\item  If one wants to add $N$ arbitrary numbers, then the number of operations required
will be $N-1$.
\item  Solving a linear system of equations $Ax=b$, being $A$ an $N\times N$ dense matrix, requires
of the order of $N^3$ operations using the typical Gaussian elimination scheme.
\item  The simplest implementations of the Hartree-Fock\index{Hartree-Fock method} method 
to find the ground state of the electronic Schr\"odinger Hamiltonian require a number of operations which
is proportional to $N^4$, being $N$ the number of basis functions used.
\item  A naive approach to calculate an estimation of the partition function 
of a system depending on $N$ coordinates,
and sampling $m$ different values for each, takes of the order of $m^N$ operations (exponential growth).
\end{itemize}
In all these examples the size of the system is proportional to a number $N$, and 
the increase of $N$ leads to
an increase of the numerical complexity of the solution of the problem. 
In doing any calculation, we want its result to be ready within a given time;
systems beyond a given size will be unaffordable.  
This scaling of the methods can sometimes be reduced by doing a number of approximations consisting of
neglecting part of the information involved in the problem and
expecting it will not have a major influence on results \cite{germann_Pflop_MD}.  

The considerations about simulation time
are more complex if parallel programs are run, instead of serial programs.
Parallel programs
distribute the workload in several computing \emph{threads}, each of which is run in a different 
computing unit.
When executing a parallel program, it is customary to measure its efficiency with
\begin{itemize}
\item  The total execution time (wall clock time\index{wall clock time})
which is required for a given task $t_{N_c}$ (which is a function of the number of cores working on it, $N_c$)
\item  The \emph{speedup}\index{speedup}, which is defined as the quotient $S_{N_c}:=t_1 / t_{N_c}$
This is, the time that the task would last if run in one core divided by the time it lasts when run in $N_c$ cores.
\item  The quotient $S_{N_c}/N_c$ (sometimes called the \emph{efficiency factor})
\end{itemize}
For a given problem of constant size, Amdahl's\index{Amdahl's Law} Law \cite{Amdahlslaw}
states that if \emph{p} is the fraction of the problem which can be run in parallel,
and therefore $s:=(1 - \emph{p})$ is the minimum fraction which must be run in serial, 
then the maximum speedup that can be achieved by using $N_c$ cores is 
\begin{equation}\label{amdahlslaw}
S_{N_c}^{max}=\frac{N_c}{N_c (1-p)+p} \ .
\end{equation}
This expression
has an horizontal asymptote in $(1-p)^{-1}$.
The speedup can be increased by increasing the total time required by the fraction
of the problem which can be run in 
parallel, which can usually be achieved by increasing the size of the simulated system (for example, 
increasing its number of atoms). Commonly, $p$ is not constant, but increasing as the size of the problem 
increases. 
Let us consider a variable-size problem which requires a time of
$T(s+pN_c^{\alpha})$ to be solved in serial. In this expression,
$T$ is the total time required for solving the problem of a given size in serial,
and the exponent $\alpha$ is a given positive number. If the part that can be parallelized is indeed parallelized
(assuming optimal scaling in the $N_c$ computing units),
the time required by the execution in parallel will be $T(s+pN_c^{\alpha-1})$. 
Applying that $s+p=1$ in the
ratio of serial and parallel times for a variable-size problem for $0<\alpha < 1$, 
becomes:
\begin{equation}\label{guga1}
S_v := \frac{s+(1-s)N_c{^\alpha}Ã}{s+(1-s)N_c^{\alpha-1}Ã}  \ , 
\end{equation}
where $S_v$ means speedup factor for variable size algorithms. 
If $\alpha=0$ (i.e., if the problem size does not increase with $N_c$) the expression above equals the
Amdahl's Law (\ref{amdahlslaw}).
In the limit of high $N_c$, eq. (\ref{guga1}) becomes 
\begin{equation}
S_v := 1 + \frac{p}{s} N^{\alpha} \ .
\end{equation}
In the case of $\alpha=1$ (linear scaling of the size of the problem with the number of computing units
solving it), the ratio of serial and parallel time is
\begin{equation}\label{guga2}
S_{v}^{G} = s + (1-s)N_c = (1-p) + p N_c \ , 
\end{equation}
for the ideal parallelization situation. Equation (\ref{guga2}) is called the
\emph{Gustafson's Law}\index{Gustafson's Law} \cite{libro_HPC} and
states that the speedup for solving a problem can be increased by increasing
the size of its parallelizable part.

Considerations such as the ones underlying Amdahl's and Gustafson's Laws can be useful
for scientific software developers\index{software developer}, in order to 
increase the efficiency of their codes. Parallelization characteristics of algorithms, however,
are commonly much harder to derive than these laws.

\section*{Acknowledgments}

The authors would like to thank Joseba Alberdi, David Strubbe, Burkhard Bunk, 
Stephen Christensen and Pablo Echenique for their illuminating help and advice. 
In addition, we would like to thank Jack Dongarra and the J\"ulich Supercomputing Center for 
the information and resources provided.

\bibliographystyle{model1-num-names}
\bibliography{refs}

\begin{thebibliography}{42}
\expandafter\ifx\csname natexlab\endcsname\relax\def\natexlab#1{#1}\fi
\providecommand{\bibinfo}[2]{#2}
\ifx\xfnm\relax \def\xfnm[#1]{\unskip,\space#1}\fi

\bibitem{makovsimulation}
{\sc G.~Makov}, {\sc C.~Gattinoni}, and {\sc A.~D. Vita},
\newblock {\em Ab initio based multiscale modelling for materials science},
\newblock Modelling and Simulation in Materials Science and Engineering {\bf
  17} (2009) 084008.


\bibitem{Cra2002Book}
{\sc C.~J. Cramer},
\newblock {\em Essentials of Computational Chemistry: Theories and Models},
\newblock John Wiley \& Sons, Chichester, 2nd edition, 2002.


\bibitem{germann_Pflop_MD}
{\sc T.~C. Germann}, {\sc K.~Kadau}, and {\sc S.~Swaminarayan},
\newblock {\em 369 Tflop/s molecular dynamics simulations on the petaflop
  hybrid supercomputer ''Roadrunner''},
\newblock Concurrency and Computation: Practice and Experience {\bf 21} (2009)
  2143--2159.


\bibitem{PhysRevLett.90.258101}
{\sc M.~A.~L. Marques}, {\sc X.~L\'opez}, {\sc D.~Varsano}, {\sc A.~Castro},
  and {\sc A.~Rubio},
\newblock {\em Time-Dependent Density-Functional Approach for Biological
  Chromophores: The Case of the Green Fluorescent Protein},
\newblock Phys. Rev. Lett. {\bf 90} (2003) 258101.


\bibitem{Shaw15102010}
{\sc D.~E. Shaw}, {\sc P.~Maragakis}, {\sc K.~Lindorff-Larsen}, {\sc S.~Piana},
  {\sc R.~O. Dror}, {\sc M.~P. Eastwood}, {\sc J.~A. Bank}, {\sc J.~M. Jumper},
  {\sc J.~K. Salmon}, {\sc Y.~Shan}, and {\sc W.~Wriggers},
\newblock {\em Atomic-Level Characterization of the Structural Dynamics of
  Proteins},
\newblock Science {\bf 330} (2010) 341-346.


\bibitem{CPHC:CPHC200600128}
{\sc D.~Marx},
\newblock {\em Proton Transfer 200 Years after von Grotthuss: Insights from Ab
  Initio Simulations},
\newblock ChemPhysChem {\bf 7} (2006) 1848--1870.


\bibitem{libro_HPC}
{\sc G.~Hager} and {\sc G.~Wellein},
\newblock {\em Introduction to High Performance Computing for Scientists and
  Engineers},
\newblock CRC Press - Taylor \& Francis Group, 1st edition, 2011.


\bibitem{MPI_OpenMP_HPC}
{\sc J.~Haoqiang}, {\sc D.~Jespersen}, {\sc P.~Mehrotra}, {\sc R.~Biswas}, {\sc
  H.~Lei}, and {\sc B.~Chapman},
\newblock {\em High performance computing using MPI and OpenMP on multi-core
  parallel systems},
\newblock Parallel Computing {\bf 37} (2011) 562--575.


\bibitem{HenYYYYBook2012}
{\sc J.~L. Hennessy} and {\sc D.~A. Patterson},
\newblock {\em Computer Architecture, A Quantitative Approach},
\newblock Morgan Kaufmann - Elsevier, San Mateo, CA, 5th edition, 2012.


\bibitem{microprocessoreport}
Microprocessor report {\bf 25,10} (2011) .


\bibitem{PARALLEL_HI-PERFORMANCE_COMPUT_CHEMISTRY}
{\sc W.~A. de~Jong}, {\sc E.~Bylaska}, {\sc N.~Govind}, {\sc C.~L. Janssen},
  {\sc K.~Kowalski}, {\sc T.~Muller}, {\sc I.~M.~B. Nielsen}, {\sc H.~J.~J. van
  Dam}, {\sc V.~Veryazov}, and {\sc R.~Lindh},
\newblock {\em Utilizing high performance computing for chemistry: parallel
  computational chemistry},
\newblock Phys. Chem. Chem. Phys. {\bf 12} (2010) 6896--6920.


\bibitem{memories}
{\sc R.~Wood},
\newblock {\em Future hard disk drive systems},
\newblock Journal of Magnetism and Magnetic Materials {\bf 321} (2009) 555 -
  561, Current Perspectives: Perpendicular Recording.


\bibitem{data_management_2005}
{\sc J.~Gray}, {\sc D.~T. Liu}, {\sc M.~Nieto-Santisteban}, {\sc A.~Szalay},
  {\sc D.~J. DeWitt}, and {\sc G.~Heber},
\newblock {\em Scientific data management in the coming decade},
\newblock SIGMOD Rec. {\bf 34} (2005) 34--41.


\bibitem{expendingmooresdividend}
{\sc J.~Larus},
\newblock {\em Spending Moore's dividend},
\newblock Commun. ACM {\bf 52} (2009) 62--69.


\bibitem{fertmagnetoresistance_original}
{\sc M.~N. Baibich}, {\sc J.~M. Broto}, {\sc A.~Fert}, {\sc F.~N. Van~Dau},
  {\sc F.~Petroff}, {\sc P.~Etienne}, {\sc G.~Creuzet}, {\sc A.~Friederich},
  and {\sc J.~Chazelas},
\newblock {\em Giant Magnetoresistance of (001)Fe/(001)Cr Magnetic
  Superlattices},
\newblock Phys. Rev. Lett. {\bf 61} (1988) 2472--2475.


\bibitem{magnetoresistance_2}
{\sc C.~Chappert}, {\sc A.~Fert}, and {\sc F.~N. Van~Dau},
\newblock {\em The emergence of spin electronics in data storage},
\newblock Nat. Mater. {\bf 6} (2007) 813--823.


\bibitem{sundongmram}
{\sc G.~Sun}, {\sc X.~Dong}, {\sc Y.~Xie}, {\sc J.~Li}, and {\sc Y.~Chen},
\newblock in {\em A novel architecture of the 3D stacked MRAM L2 cache for CMPs
  in IEEE 15th International Symposium on High Performance Computer
  Architecture}, pp. 239--249, IEEE, 2009.

\bibitem{storage_1}
{\sc Y.~Jin} and {\sc K.~Li},
\newblock {\em An optimal multimedia object allocation solution in
  multi-powermode storage systems},
\newblock Concurrency and Computation: Practice and Experience {\bf 22} (2010)
  1852--1873.


\bibitem{vonneumanoriginal}
{\sc J.~von Neumann},
\newblock {\em First draft of a report on the EDVAC},
\newblock Annals of the History of Computing, IEEE {\bf 15} (1993 (original
  from 1945)) 27--75.


\bibitem{future_of_microprocessors}
{\sc K.~Olukotun} and {\sc L.~Hammond},
\newblock {\em The Future of Microprocessors},
\newblock Queue {\bf 3} (2005) 26--29.


\bibitem{Moore1965law}
{\sc G.~Moore},
\newblock {\em Cramming more components onto integrated circuits},
\newblock Electronics {\bf 38} (1965) 56--59.


\bibitem{multicore_math}
{\sc A.~Buttari}, {\sc J.~Dongarra}, {\sc J.~Kurzak}, {\sc J.~Langou}, {\sc
  P.~Luszczek}, and {\sc S.~Tomov},
\newblock {\em The impact of multicore on math software},
\newblock in {\em Proceedings of the 8th international conference on Applied
  parallel computing: state of the art in scientific computing}, PARA'06, pp.
  1--10, Berlin, Heidelberg, 2007, Springer-Verlag.

\bibitem{Cas2006PSS}
{\sc A.~Castro}, {\sc H.~Appel}, {\sc M.~Oliveira}, {\sc C.~A. Rozzi}, {\sc
  X.~Andrade}, {\sc F.~Lorenzen}, {\sc M.~A.~L. Marques}, {\sc E.~K.~U. Gross},
  and {\sc A.~Rubio},
\newblock {\em Octopus: a tool for the application of time-dependent density
  functional theory},
\newblock Phys. Stat. Sol {\bf 243} (2006) 2465.


\bibitem{Mar2003CPC}
{\sc M.~A.~L. Marques}, {\sc A.~Castro}, {\sc G.~F. Bertsch}, and {\sc
  A.~Rubio},
\newblock {\em octopus: a first-principles tool for excited electron-ion
  dynamics},
\newblock Computer Physics Communications {\bf 151} (2003) 60.


\bibitem{octopuslessi}
{\sc X.~Andrade}, {\sc J.~Alberdi-Rodriguez}, {\sc D.~A. Strubbe}, {\sc
  M.~J.~T. Oliveira}, {\sc F.~Nogueira}, {\sc A.~Castro}, {\sc J.~Muguerza},
  {\sc A.~Arruabarrena}, {\sc S.~G. Louie}, {\sc A.~Aspuru-Guzik}, {\sc
  A.~Rubio}, and {\sc M.~A.~L. Marques},
\newblock {\em TDDFT in massively parallel computer architectures: The OCTOPUS
  project},
\newblock Psi-k Scientific Highlights (highlight of the month) {\bf April}
  (2012) .


\bibitem{IBM_meteorology}
{\sc S.~Zhou}, {\sc D.~Duffy}, {\sc T.~Clune}, {\sc M.~Suarez}, {\sc
  S.~Williams}, and {\sc M.~Halem},
\newblock {\em The impact of IBM Cell technology on the programming paradigm in
  the context of computer systems for climate and weather models},
\newblock Concurrency and Computation: Practice and Experience {\bf 21} (2009)
  2176--2186.


\bibitem{Latboltz}
{\sc S.~Williams}, {\sc J.~Carter}, {\sc L.~Oliker}, {\sc J.~Shalf}, and {\sc
  K.~Yelick},
\newblock {\em Lattice Boltzmann simulation optimization on leading multicore
  platforms},
\newblock Proceedings of the IEEE International Symposium on Parallel and
  Distributed Processing, 2008.  (2008) 1--14.


\bibitem{simdastronomia}
{\sc A.~Tanikawa}, {\sc K.~Yoshikawa}, {\sc T.~Okamoto}, and {\sc K.~Nitadori},
\newblock {\em N-body simulation for self-gravitating collisional systems with
  a new SIMD instruction set extension to the x86 architecture, Advanced Vector
  eXtensions},
\newblock New Astronomy {\bf 17} (2011) 82--92.


\bibitem{gpuhouston}
{\sc J.~D. Owens}, {\sc M.~Houston}, {\sc D.~Luebke}, {\sc S.~Green}, {\sc
  J.~E. Stone}, and {\sc J.~C. Phillips},
\newblock {\em GPU computing},
\newblock Proceedings of the IEEE {\bf 96} (2008) 879--899.


\bibitem{gromacs2001}
{\sc E.~Lindahl}, {\sc B.~Hess}, and {\sc D.~van~der Spoel},
\newblock {\em GROMACS 3.0: a package for molecular simulation and trajectory
  analysis},
\newblock Journal of Molecular Modeling {\bf 7} (2001) 306--317.


\bibitem{intelarch2011}
{\sc Intel},
\newblock {\em Intel 64 and IA-32 Architectures Software Developer's Manual},
  2011,
\newblock Intel order Number: 325462-039US.

\bibitem{foldingathomeresults}
{\sc V.~S. Pande}, {\sc I.~Baker}, {\sc J.~Chapman}, {\sc S.~P. Elmer}, {\sc
  S.~Khaliq}, {\sc S.~M. Larson}, {\sc Y.~M. Rhee}, {\sc M.~R. Shirts}, {\sc
  C.~D. Snow}, {\sc E.~J. Sorin}, and {\sc B.~Zagrovic},
\newblock {\em Atomistic protein folding simulations on the submillisecond time
  scale using worldwide distributed computing},
\newblock Biopolymers {\bf 68} (2003) 91--109.


\bibitem{HenYYYYBookappI}
{\sc J.~L. Hennessy} and {\sc D.~A. Patterson},
\newblock {\em Appendix H: Large-Scale Multiprocessors and Scientific
  Applications},
\newblock in {\em Computer Architecture, A Quantitative Approach}, Morgan
  Kaufmann - Elsevier, San Mateo, CA, 4th edition, 2007.

\bibitem{bluegenebook}
{\sc C.~Sosa} and {\sc B.~Knudson},
\newblock {\em IBM System Blue Gene Solution: Blue Gene/P Application
  Development},
\newblock IBM Redbooks, 2009.


\bibitem{guialargampi}
{\sc J.~Dongarra}, {\sc D.~Walker}, {\sc E.~Lusk}, {\sc M.~Snir}, {\sc
  W.~Gropp}, {\sc A.~Geist}, {\sc S.~Otto}, {\sc R.~Hempel}, {\sc J.~Cownie},
  {\sc T.~Skjelum}, {\sc L.~Clarke}, {\sc R.~Littlefield}, {\sc M.~Sears}, and
  {\sc S.~Huss-Lederman},
\newblock {\em MPI: A Message-Passing Interface Standard},
\newblock University of Tennessee, Knoxville, Tennessee, 4th edition, 2003.


\bibitem{hetero_chips}
{\sc R.~Kumar}, {\sc D.~Tullsen}, {\sc N.~Jouppi}, and {\sc P.~Ranganathan},
\newblock {\em Heterogeneous chip multiprocessors},
\newblock Computer {\bf 38} (2005) 32--38.


\bibitem{hi-performance_comp_gpu_nvidia}
{\sc E.~Lindholm}, {\sc J.~Nickolls}, {\sc S.~Oberman}, and {\sc J.~Montrym},
\newblock {\em NVIDIA Tesla: A Unified Graphics and Computing Architecture},
\newblock Micro, IEEE {\bf 28} (2008) 39 -55.


\bibitem{Dro2010JCB}
{\sc R.~O. Dror}, {\sc M.~O. Jensen}, {\sc D.~W. Borhani}, and {\sc D.~E.
  Shaw},
\newblock {\em Exploring atomic resolution physiology on a femtosecond to
  millisecond timescale using molecular dynamics simulations},
\newblock The Journal of Cell Biology {\bf 135 (6)} (2010) 555--562.


\bibitem{Sha2009XXX}
{\sc D.~E. Shaw}, {\sc R.~Ron O.~Dror}, {\sc J.~Salmon}, {\sc J.~Grossman},
  {\sc K.~Mackenzie}, {\sc J.~Bank}, {\sc C.~Young}, {\sc B.~Batson}, {\sc
  K.~Bowers}, {\sc E.~Edmond~Chow}, {\sc M.~Eastwood}, {\sc D.~Ierardi}, {\sc
  J.~John L.~Klepeis}, {\sc J.~Jeffrey S.~Kuskin}, {\sc R.~Larson}, {\sc
  K.~Kresten Lindorff-Larsen}, {\sc P.~Maragakis}, {\sc M.~M.A.}, {\sc
  S.~Piana}, {\sc S.~Yibing}, and {\sc B.~Towles},
\newblock {\em Millisecond-Scale Molecular Dynamics Simulations on Anton},
\newblock In Proceedings of the ACM/IEEE Conference on Supercomputing (SC09),
  ACM Press, New York,  (2009) .


\bibitem{special_purpose_janus}
{\sc F.~Belletti}, {\sc M.~Cotallo}, {\sc A.~Cruz}, {\sc L.~A. Fernandez}, {\sc
  A.~Gordillo-Guerrero}, {\sc M.~Guidetti}, {\sc A.~Maiorano}, {\sc
  F.~Mantovani}, {\sc E.~Marinari}, {\sc V.~Martin-Mayor}, {\sc
  A.~Munoz-Sudupe}, {\sc D.~Navarro}, {\sc G.~Parisi}, {\sc S.~Perez-Gaviro},
  {\sc M.~Rossi}, {\sc J.~J. Ruiz-Lorenzo}, {\sc S.~F. Schifano}, {\sc
  D.~Sciretti}, {\sc A.~Tarancon}, {\sc R.~Tripiccione}, {\sc J.~L. Velasco},
  {\sc D.~Yllanes}, and {\sc G.~Zanier},
\newblock {\em Janus: An FPGA-Based System for High-Performance Scientific
  Computing},
\newblock Computing in Science and Engineering {\bf 11} (2009) 48-58.


\bibitem{special-purpose_astronomy}
{\sc J.~Makino}, {\sc T.~Fukushige}, {\sc M.~Koga}, and {\sc K.~Namura},
\newblock {\em GRAPE-6: The massively-parallel special-purpose computer for
  astrophysical particle simulations},
\newblock Publ. Astron. Soc. Japan {\bf 55} (2003) 1163.


\bibitem{linpackb2003}
{\sc J.~J. Dongarra}, {\sc P.~Luszczek}, and {\sc A.~Petitet},
\newblock {\em The LINPACK benchmark: Past, present and future},
\newblock Concurrency and Computation: Practice and Experience {\bf 15} (2003)
  803--820.


\bibitem{profiling_mohr}
{\sc B.~Mohr}, {\sc B.~J.~N. Wylie}, and {\sc F.~Wolf},
\newblock {\em Performance measurement and analysis tools for extremely
  scalable systems},
\newblock Concurrency and computation: Practice and experience {\bf 22} (2010)
  2212--2229.


\bibitem{energy_saving_2}
{\sc D.~H. Woo} and {\sc H.-H. Lee},
\newblock {\em Extending Amdahl's Law for Energy-Efficient Computing in the
  Many-Core Era},
\newblock Computer {\bf 41} (2008) 24 -31.


\bibitem{Hamilton_Amazon}
{\sc J.~Hamilton},
\newblock {\em CooperativeExpendableMicro-SliceServers (CEMS): Low Cost, Low
  Power Servers for Internet-Scale Services},
\newblock Proc. 4th Biennial Conf. Innovative Data Systems Research (CIDR),
  Asilomar, CA, USA, January., 2009.


\bibitem{energy_consumption}
{\sc L.~Barroso} and {\sc U.~Holzle},
\newblock {\em The Case for Energy-Proportional Computing},
\newblock Computer {\bf 40} (2007) 33 -37.


\bibitem{energy_saving_cloud}
{\sc A.~Berl}, {\sc E.~Gelenbe}, {\sc M.~d. Girolamo}, {\sc G.~Giuliani}, {\sc
  H.~De~Meer}, {\sc Q.~D. Mihn}, and {\sc K.~Pentikousis},
\newblock {\em Energy-Efficient Cloud Computing},
\newblock The Computer Journal {\bf 53} (2010) 1045--1051.


\bibitem{libro_grid_computing_ch1}
{\sc B.~Wilkinson},
\newblock {\em Grid Computing: Techniques and Applications},
\newblock Chapman \& Hall/CRC Computational Science, 1st edition, 2009.


\bibitem{gridDNA}
{\sc M.-A. Thyveetil}, {\sc P.~V. Coveney}, {\sc H.~C. Greenwell}, and {\sc
  J.~L. Suter},
\newblock {\em Computer Simulation Study of the Structural Stability and
  Materials Properties of DNA-Intercalated Layered Double Hydroxides},
\newblock J. Am. Chem. Soc. {\bf 130} (2008) 4742-4756.


\bibitem{grid_kondo}
{\sc D.~Kondo}, {\sc B.~Javadi}, {\sc P.~Malecot}, {\sc F.~Cappello}, and {\sc
  D.~P. Anderson},
\newblock {\em Cost-benefit analysis of Cloud Computing versus desktop grids},
\newblock Parallel and Distributed Processing Symposium, International {\bf 0}
  (2009) 1-12.


\bibitem{foldingathomepaper}
{\sc A.~L. Beberg}, {\sc D.~L. Ensign}, {\sc G.~Jayachandran}, {\sc S.~Khaliq},
  and {\sc V.~S. Pande},
\newblock {\em Folding@home: Lessons from eight years of volunteer distributed
  computing},
\newblock Parallel and Distributed Processing Symposium, International {\bf 0}
  (2009) 1-8.


\bibitem{cloud_computing_astronomy}
{\sc C.~Hoffa}, {\sc G.~Mehta}, {\sc T.~Freeman}, {\sc E.~Deelman}, {\sc
  K.~Keahey}, {\sc B.~Berriman}, and {\sc J.~Good},
\newblock {\em On the Use of Cloud Computing for Scientific Workflows},
\newblock in {\em eScience, 2008. eScience '08. IEEE Fourth International
  Conference on}, pp. 640 --645, 2008.

\bibitem{cloud_foster}
{\sc I.~Foster}, {\sc Y.~Zhao}, {\sc I.~Raicu}, and {\sc S.~Lu},
\newblock {\em Cloud Computing and Grid Computing 360-Degree Compared},
\newblock Grid Computing Environments Workshop, 2008. GCE '08  (2009) 1--10.


\bibitem{HenYYYYBook}
{\sc D.~Goldberg},
\newblock {\em Appendix H: Computer Arithmetic},
\newblock in {\em Computer Architecture, A Quantitative Approach}, edited by
  {\sc J.~L. Hennessy} and {\sc D.~A. Patterson}, Morgan Kaufmann - Elsevier,
  San Mateo, CA, 2003.

\bibitem{softerrors}
{\sc P.~Shivakumar}, {\sc M.~Kistler}, {\sc S.~Keckler}, {\sc D.~Burger}, and
  {\sc L.~Alvisi},
\newblock {\em Modeling the effect of technology trends on the soft error rate
  of combinational logic},
\newblock in {\em Dependable Systems and Networks, 2002. DSN 2002. Proceedings.
  International Conference on}, pp. 389--398, 2002.

\bibitem{Amdahlslaw}
{\sc G.~M. Amdahl},
\newblock {\em Validity of the single processor approach to achieving large
  scale computing capabilities},
\newblock in {\em Proceedings of the April 18-20, 1967, spring joint computer
  conference}, AFIPS '67 (Spring), pp. 483--485, New York, NY, USA, 1967, ACM.

\end{thebibliography}

\end{document}